\documentclass[nofootinbib,amsmath,notitlepage,preprintnumbers]{revtex4-1}
\usepackage{multirow}
\usepackage{amssymb,esvect,amsmath,graphicx,latexsym,amsthm,slashed,eso-pic}
             \DeclareMathOperator{\s}{s}

\newcommand{\ie}{{\it i.e.}}

\newcommand{\beq}{\begin{equation}} \newcommand{\eeq}{\end{equation}}
\newcommand{\bea}{\begin{eqnarray}} \newcommand{\eea}{\end{eqnarray}}

\newcommand{\alterbbn}{\texttt{AlterBBN}}

\newcommand{\lsim}{\mathrel{\hbox{\rlap{\lower.55ex\hbox{$\sim$}} \kern-.3em \raise.4ex \hbox{$<$}}}}
\newcommand{\gsim}{\mathrel{\hbox{\rlap{\lower.55ex\hbox{$\sim$}} \kern-.3em \raise.4ex \hbox{$>$}}}}

\def\lsim{\mathrel{\raise.3ex\hbox{$<$\kern-.75em\lower1ex\hbox{$\sim$}}}}
\def\gsim{\mathrel{\raise.3ex\hbox{$>$\kern-.75em\lower1ex\hbox{$\sim$}}}}

\newcommand{\be}{\begin{eqnarray}}
\newcommand{\ee}{\end{eqnarray}}

\newcommand{\benum}{\begin{enumerate}}
\newcommand{\eenum}{\end{enumerate}}
\newcommand{\bi}{\begin{itemize}}
\newcommand{\ei}{\end{itemize}}

\begin{document}

\preprint{FERMILAB-PUB-20-224-A}

\title{Constraints on Primordial Black Holes From Big Bang Nucleosynthesis Revisited}

\author{Celeste Keith$^{a,b}$}
\thanks{ORCID: https://orcid.org/0000-0002-3004-0930}

\author{Dan Hooper$^{a,b,c}$}
\thanks{ORCID: http://orcid.org/0000-0001-8837-4127}

\author{Nikita Blinov$^c$}
\thanks{ORCID: http://orcid.org/0000-0002-2845-961X}

\author{Samuel D.~McDermott$^{c}$}
\thanks{ORCID: http://orcid.org/0000-0001-5513-1938}

\affiliation{$^a$University of Chicago, Kavli Institute for Cosmological Physics}
\affiliation{$^b$University of Chicago, Department of Astronomy and Astrophysics}
\affiliation{$^c$Fermi National Accelerator Laboratory, Theoretical Astrophysics Group}

\date{\today}

\begin{abstract}

As space expands, the energy density in black holes increases relative to that of radiation, providing us with motivation to consider scenarios in which the early universe contained a significant abundance of such objects. In this study, we revisit the constraints on primordial black holes derived from measurements of the light element abundances. Black holes and their Hawking evaporation products can impact the era of Big Bang Nucleosynthesis (BBN) by altering the rate of expansion at the time of neutron-proton freeze-out, as well as by radiating mesons which can convert protons into neutrons and vice versa. Such black holes can thus enhance the primordial neutron-to-proton ratio, and increase the amount of helium that is ultimately produced. Additionally, the products of Hawking evaporation can break up helium nuclei, which both reduces the helium abundance and increases the abundance of primordial deuterium. Building upon previous work, we make use of modern deuterium and helium measurements to derive stringent constraints on black holes which evaporate in $t_{\rm evap} \sim 10^{-1}$ s to $\sim 10^{13}$ s (corresponding to $M \sim 6\times 10^8$ g to $\sim 2 \times 10^{13}$ g, assuming Standard Model particle content). We also consider how physics beyond the Standard Model could impact these constraints. Due to the gravitational nature of Hawking evaporation, the rate at which a black hole evaporates, and the types of particles that are produced through this process, depend on the complete particle spectrum. Within this context, we discuss scenarios which feature a large number of decoupled degrees-of-freedom (\ie~large hidden sectors), as well as models of TeV-scale supersymmetry.

\end{abstract}

\maketitle

\section{Introduction}

Although our universe is approximately homogeneous on the scales probed by the cosmic microwave background (CMB) and large scale structure, it is possible that large inhomogeneities could exist on smaller scales. If the amplitude of such inhomogeneities is sufficiently large, these density perturbations could have lead to the formation of primordial black holes in the early universe~\cite{Carr:1974nx,Carr:1975qj}. Alternatively, significant abundances of black holes could have been formed as a result of phase transitions prior to BBN~\cite{Hawking:1982ga,Sasaki:1982fi,Lewicki:2019gmv}.

Although the mass of the black holes that may have formed in the early universe is model dependent, we might reasonably expect this mass to be comparable to the total energy enclosed within the horizon at the time of their formation. In a radiation dominated era, this mass is given by~\cite{GarciaBellido:1996qt,Kawasaki:2016pql,Clesse:2016vqa,Kannike:2017bxn,Kawasaki:1997ju,Cai:2018rqf,Yoo:2018esr,Young:2015kda,Clesse:2015wea,Hsu:1990fg,La:1989za,La:1989st,La:1989pn,Weinberg:1989mp,Steinhardt:1990zx,Accetta:1989cr,Holman:1990wq,Hawking:1982ga,Khlopov:1980mg}:
\begin{equation}
\label{horizon}
M_{\rm hor} = \frac{M^2_{\rm Pl}}{2H} \sim 10^{10} \, {\rm g} \times \bigg(\frac{10^{11} \, {\rm GeV}}{T}\bigg)^2 \, \bigg(\frac{106.75}{g_{\star}(T)}\bigg)^{1/2},
\end{equation}
where $M_{\rm Pl} = 1.22 \times 10^{19}$ GeV is the Planck mass, $H$ is the Hubble rate, $T$ is the temperature of radiation, and $g_{\star}(T)$ is effective number of relativistic degrees-of-freedom. This provides us with motivation to consider a wide range of masses, extending from very small black holes (which evaporate well before the onset of BBN), to black holes with masses as large as $M \sim 10^2 M_{\odot}$, which may have formed shortly before the BBN era.

If there existed even a very small abundance of black holes in the early universe, they would make up an increasingly large fraction of the total energy density as the universe expands, with the ratio $\rho_{\rm BH}/\rho_{\rm rad}$ growing proportionally to the scale factor during the era of radiation domination (see, for example, Refs.~\cite{Lennon:2017tqq,Morrison:2018xla,Hooper:2019gtx}). If the black holes are very massive and long-lived, they could make up all or some of the dark matter in the universe today (see, for example, Refs.~\cite{Bird:2016dcv,Carr:2016drx}). Alternatively, there may have been an era prior to BBN in which much smaller black holes dominated the total energy density, up to the point of their evaporation and the subsequent reheating of the radiation bath through Hawking radiation.

Once black holes have formed, they lose mass through the process of Hawking evaporation. For the case of a Schwarzschild black hole of mass $M$, the rate for this process is given by~\cite{Hawking:1974sw}:
\begin{eqnarray}
\label{loss}
\frac{dM}{dt} = -\frac{\mathcal{G} g_{\star, H}(T_{\rm BH}) M^4_{\rm Pl}}{30720 \pi M^2} \simeq -8.2 \times 10^{6} \, {\rm g/s} \,\bigg(\frac{g_{\star, H}}{108}\bigg) \, \bigg(\frac{10^{10} \, {\rm g}}{M}\bigg)^2,
\end{eqnarray}
where $\mathcal{G} \approx 3.8$ is the appropriate greybody factor and the temperature of a black hole is related to its mass as follows:
\begin{eqnarray}
T_{\rm BH} = \frac{M^2_{\rm Pl}}{8\pi M} \simeq 1.05 \, {\rm TeV} \times \bigg(\frac{10^{10} \, {\rm g}}{M}\bigg).
\end{eqnarray}
The quantity $g_{\star, H}(T_{\rm BH})$ counts the number of particle degrees-of-freedom with masses below $\sim$\,$T_{\rm BH}$, weighted by factors of approximately 1.82, 1.0, 0.41 or 0.05 for particles of spin $0$, $1/2$, $1$ or $2$, respectively~\cite{1990PhRvD..41.3052M,1991PhRvD..44..376M}. For $T_{\rm BH} \gg 100$ GeV ($M_{\rm BH} \ll 10^{11}$ g), the particle content of the Standard Model corresponds to $g_{\star, H} \simeq 108$. For Kerr black holes (\ie~black holes with appreciable angular momentum), the values of $g_{\star, H}$ are somewhat higher, and favor the radiation of high-spin particles~\cite{Page:1976ki,Taylor:1998dk,Hooper:2020evu}. In this study, we will limit ourselves to the case of non-rotating (\ie~Schwarzschild) black holes.

As a black hole loses mass, it emits Hawking radiation at a greater rate, causing it to evaporate over the following timescale:
\begin{eqnarray}
\label{tevap}
t_{\rm evap} = \frac{30720\pi}{\mathcal{G}M^4_{\rm Pl}} \int^{M_i}_0 \frac{dM_{\rm BH} M_{\rm BH}^2}{g_{\star, H}(T_{\rm BH})} \approx 4.0 \times 10^{2} \, {\rm s} \times \bigg(\frac{M_i}{10^{10} \, {\rm g}}\bigg)^3 \bigg(\frac{108}{\langle g_{\star, H}\rangle}\bigg), 
\end{eqnarray}
where $M_i$ is the initial mass of the black hole and $\langle g_{\star, H} \rangle^{-1} \equiv (3/M^3_i) \int^{M_i}_0 dM M^2/g_{\star, H}(T_{\rm BH})$ is the value of $g_{\star, H}$ appropriately averaged over the course of the black hole's evaporation. For black holes with evaporation times between $t_{\rm evap} \sim 10^{-1}$ s and $\sim 10^{13}$ s (corresponding to $M \sim 6\times 10^8$ g to $\sim 2 \times 10^{13}$ g, assuming Standard Model particle content), measurements of the light element abundances typically provide us with the most stringent constraints on their abundance. For longer evaporation times, measurements of the CMB are generally more restrictive. For a review of constraints on primordial black holes, see Ref.~\cite{Carr:2020gox}. 

In this study, we revisit the constraints on primordial black holes that can be derived from measurements of the primordial light element abundances. In particular, we use modern measurements of primordial hydrogen, deuterium and helium to derive upper limits on the initial abundances of $M\sim10^8-10^{13} \, {\rm g}$ black holes. For black holes heavier than $\sim 10^{10} \, {\rm g}$, the strongest constraints result from the photodissociation or hadrodissociation of helium nuclei and the corresponding production of antideuterons. Lighter black holes are constrained by their impact on the Hubble rate, which can alter the time at which the weak interactions effectively freeze-out, as well as the Hawking radiation of hadrons and mesons, each of which can alter the neutron-to-proton ratio and enhance the resulting helium abundance. We also consider how these constraints can change in the presence of particle content beyond the Standard Model. The existence of additional particle species can increase the rate at which black holes evaporate, typically weakening the resulting constraints. Furthermore, in scenarios that feature large numbers of decoupled degrees-of-freedom, the fraction of a black hole's mass that goes into particles that can break up helium can be significantly reduced. If stable, such Hawking evaporation products can act as dark radiation or dark matter.

\section{Measurements of the Primordial Light Element Abundances}

In this study, we make use of two sets of measurements of the primordial light element abundances:
\begin{itemize}
\item{We use the deuterium-to-hydrogen measurement of $({\rm D}/{\rm H})_p = (2.53 \pm 0.04) \times 10^{-5}$, based on the observation of four damped Lyman-alpha systems~\cite{Cooke:2013cba} (see also, Ref.~\cite{Riemer-Sorensen:2017vxj}). Note that the uncertainties quoted for this result are significantly smaller than those associated with previous measurements, allowing us to place constraints that are significantly more stringent than those presented in Refs.~\cite{Carr:2009jm,Kohri:1999ex}.}
\item{For the helium mass fraction, we adopt $Y_p \equiv \rho(^4{\rm He})/\rho_b = 0.2449 \pm 0.0040$, based on the measurements of recombination lines emitted from 45 extragalactic HII regions~\cite{Izotov:2014fga}, as statistically analyzed in Ref.~\cite{Aver:2015iza}. While other recent determinations are not in total agreement (including $Y_p = 0.2446 \pm 0.0029$~\cite{Peimbert:2016bdg} and $Y_p = 0.2551 \pm 0.0022$~\cite{Izotov:2014fga}), the measurement adopted in this study is consistent with (and has slightly larger error bars than) that recommended in the Particle Data Group's BBN review~\cite{Tanabashi:2018oca}.}
\end{itemize}

For two reasons, we do not explicitly make use of $^3$He measurements in this study. First, measurements of primordial $^3$He are complicated by the fact that stellar nucleosynthesis models for $^3$He are in conflict with observations~\cite{Olive:1996tt}. In light of this, it may be unwise to treat $^3$He as a reliable probe of the early universe~\cite{Tanabashi:2018oca}. Second, in light of recent improvements in the precision of primordial deuterium measurements, the constraints one might derive from $^3$He are most stringent only in small corners of parameter space (for example, in a scenario in which black holes evaporate $\sim10^6-10^8$ s after the Big Bang to a large number of electromagnetically-charged degrees-of-freedom, beyond those of the Standard Model). On similar grounds~\cite{Tanabashi:2018oca}, we do not make use of primordial lithium measurements in our analysis.

\section{The Impact of Evaporating Black Holes on Big Bang Nucleosynthesis}

A great deal of effort has been invested in developing sophisticated codes which can make detailed and accurate predictions for the primordial element abundances. These predictions have been found to be in excellent agreement with the measured abundances of primordial D, $^3$He and $^4$He, demonstrating that our universe was radiation dominated and generally well-described by the standard $\Lambda$CDM cosmological model during the era of primordial nucleosynthesis~\cite{Schramm:1997vs,Steigman:2007xt,Iocco:2008va,Cyburt:2015mya,Pitrou:2018cgg,Fields:2019pfx}.\footnote{The measured lithium abundance is somewhat higher than predicted by standard BBN models~\cite{Fields:2011zzb}. At this time, it is not clear whether this is a consequence of new physics, or the result of challenges associated with accurately measuring the primordial abundance of this nuclear species.} These measurements can be used to place stringent constraints on the expansion history of the universe as early as a few seconds after the Big Bang, as well as on any energy injection that may have occurred during or after this era~\cite{Sarkar:1995dd,Jedamzik:2009uy,Pospelov:2010hj,Hufnagel:2017dgo,Huang:2017egl,Forestell:2018txr,Hufnagel:2018bjp,Depta:2019lbe,Carr:2009jm,Kawasaki:2017bqm}.

The evaporation of primordial black holes could potentially impact the resulting light element abundances in a number of different ways. In this discussion, we will focus on the following four mechanisms, each of which can play a significant role: 
\begin{itemize}
\item{At a temperature near $\sim 1$ MeV, the rate of weak interactions (which can convert neutrons into protons and vice versa) falls below the rate of Hubble expansion, freezing-in the value of the neutron-to-proton ratio. The presence of black holes and their evaporation products can increase the expansion rate during this era, causing these interactions to freeze-out earlier, enhancing the neutron-to-proton ratio during BBN, and increasing the amount of helium that is ultimately produced.}
\item{Hadrons and mesons radiated from black holes can alter the neutron-to-proton ratio after the weak interactions have frozen out through processes such as $n + \pi^+ \leftrightarrow p + \pi^0$ and \,$p+\pi^-  \leftrightarrow n + \pi^0$. This can enhance the neutron-to-proton ratio during BBN and increase the resulting helium abundance.}
\item{Energetic photons from black holes can break up helium nuclei through photodissociation, reducing the resulting helium abundance and (more importantly) increasing the abundance of primordial deuterium. This process is effective, however, only if the temperature of the background radiation is too low to absorb the dissociating photons through $e^+ e^-$ pair production, $T \lsim m^2_e/22 E_{\rm He} \sim 0.4$ keV (where $E_{\rm He} \simeq 28.3$ MeV is the nuclear binding energy of helium)~\cite{Kawasaki:1994af}.}
\item{At earlier times ($T \gsim 0.4$ keV), energetic photons are typically absorbed before they can break up any helium nuclei. During this era, helium nuclei are most efficiently broken up by the energetic mesons that are radiated from black holes (\ie \,hadrodissociation).}
\end{itemize}

These and other processes have been modeled in detail by a number of modern BBN codes~\cite{Arbey:2018zfh,Arbey:2011nf,Pitrou:2019nub,Pisanti:2007hk,Coc:2011az,Coc:2017pxv}, and the impact of evaporating black holes on these processes has been studied in the past~\cite{Carr:2009jm,Kohri:1999ex,Acharya:2020jbv}. In particular, Carr {\it et al.} (2010)~\cite{Carr:2009jm,Carr:2020gox} used primordial measurements of $Y_p$, D/H, $^3$He/D and $^6$Li/$^7$Li to constrain the abundance of primordial black holes with evaporation times in the range of  $t_{\rm evap} \sim 1-10^{13} \, {\rm s}$. Although that study considered a wide range of hadronic and electromagnetic interactions, significant progress has been made in the past decade in improving these measurements, as well as in refining the codes that calculate the resulting light element abundances. Furthermore, these previous studies did not consider how the existence of particle content beyond the Standard Model could potentially alter these constraints. 

In this study, we revisit the impact of evaporating black holes on the formation of primordial nuclei, making use of the recent study by Kawasaki {\it et al.} (2018), who have used a sophisticated code to study the effects of long-lived particles on BBN~\cite{Kawasaki:2017bqm}.\footnote{For earlier work studying the impact of long-lived decaying particles on BBN, see Refs.~\cite{Forestell:2018txr,Poulin:2015opa,Cyburt:2009pg,Kusakabe:2008kf,Jedamzik:2006xz,Kohri:2005wn,Jedamzik:2004er,Kawasaki:2004qu,Kawasaki:2004yh,Kawasaki:2004qu,Kawasaki:2004qu,Cyburt:2002uv,Kohri:2001jx,Kawasaki:2000qr,Holtmann:1998gd,Kawasaki:1994sc}.} In many respects, evaporating black holes alter the predicted light element abundances in ways that are similar to decaying particles. That being said, decaying particles and evaporating black holes typically produce particles in different ratios (\ie~branching fractions), with a different distribution of energies, and with a different time profile. In what follows, we will describe our procedure for adapting constraints on long-lived decaying particles to the case of evaporating black holes.

In Kawasaki {\it et al.}~\cite{Kawasaki:2017bqm}, the authors present their results in terms of the decaying particle mass multiplied by the number of such particles per unit entropy, $M Y$, as evaluated at $t \ll \tau_X$. In contrast, Carr {\it et al.}~\cite{Carr:2009jm,Carr:2020gox} present their constraints on evaporating black holes in terms of the quantity $\beta'$, which is closely related to $\beta \equiv \rho_{\rm BH}/\rho$ evaluated at the time of black hole formation, $t_{\rm form}$. Through the following, $\beta$ can be directly related to the quantity constrained in Ref.~\cite{Kawasaki:2017bqm}, $M Y$:
\begin{eqnarray}
\beta \equiv \frac{\rho_{\rm BH}(t_{\rm form})}{\rho(t_{\rm form})} =  \frac{M n_{\rm BH}(t_{\rm form})}{\pi^2 g_{\star}(T_{\rm form})T^4_{\rm form}/30} = \frac{4}{3}\frac{M}{T_{\rm form}} \bigg(\frac{n_{\rm BH}}{s}\bigg) \equiv \frac{4}{3}\frac{MY}{T_{\rm form}},
\end{eqnarray}
where $T_{\rm form}$ is the temperature at the time of formation for a black hole of mass $M$. Carr {\it et al.} further introduced the quantity $\gamma$, which is the mass of the black hole divided by the mass enclosed within the horizon (see Eq.~\ref{horizon}) at the time of formation. This allows us to write the formation temperature as $T_{\rm form} = (45 \gamma^2 M_{\rm Pl}^6/16\pi^3 g_{\star}(T_{\rm form})M^2)^{1/4}$, and to express $\beta$ as follows:
\begin{eqnarray}
\beta = \frac{4}{3} \bigg(\frac{16\pi^3 g_{\star}(T_{\rm form})M^2}{45 \gamma^2 M_{\rm Pl}^6}\bigg)^{1/4} \, M Y.
\end{eqnarray}

For convenience, Carr {\it et al.} introduces the quantity $\beta'$, which is $\beta$ multiplied by the following powers of $\gamma$ and $g_{\star}$:
\begin{eqnarray}
\label{carrkawasaki}
\beta' \equiv \gamma^{1/2} \bigg(\frac{106.75}{g_{\star}}\bigg)^{1/4} \, \beta =   \frac{4}{3} \bigg(\frac{16\pi^3M^2 \times 106.75}{45 M_{\rm Pl}^6}\bigg)^{1/4} \, M Y.
\end{eqnarray}
This expression directly relates the ways in which Carr {\it et al.} and Kawasaki {\it et al.} characterize the magnitude of energy injection, allowing us to convert between these quantities. To put this in terms that the reader may find more intuitive, we can also relate $\beta'$ to $\Omega_{\rm BH}$, which we define as the value of $\rho_{\rm BH}/\rho_{\rm crit}$  that would be the case today if the black holes had not evaporated:
\begin{equation}
\label{omegabh}
\beta' \simeq  2.2 \times 10^{-20} \times  \bigg(\frac{\Omega_{\rm BH}}{1}\bigg) \, \bigg(\frac{M}{10^{10}\,{\rm g}}\bigg)^{0.5}.
\end{equation}
Alternatively, we can write $\beta'$ in terms of the ratio of densities in black holes and matter (at $t \ll t_{\rm evap}$):
\begin{equation}
\beta'  \simeq 7.0 \times 10^{-21} \times  \bigg(\frac{M}{10^{10} \, {\rm g}}\bigg)^{0.5} \, \frac{\rho_{\rm BH}}{\rho_M}\bigg|_{t\ll t_{\rm evap}}.
\end{equation}

While the branching fractions for the decays of a generic long-lived particle are entirely model dependent, Hawking evaporation produces various particle species with a calculable ratio, proportional to $g_{\star, H}$ (as introduced in Eq.~\ref{loss}). In contrast to decaying particles, Hawking evaporation is a purely gravitational phenomenon, and thus does not depend on the charges or interactions of the particles being radiated. For the case of Standard Model particle content, and for black holes in the mass range under consideration in this study, approximately 73\% of the total energy radiated from a black hole is in the form of quarks and gluons (and 94.5\% of the energy goes into particles other than neutrinos). When translating limits for decaying particles, we thus reduce the total decay rate by these factors (depending on whether we are in the hadrodissociation or photodissociation limit, respectively).

\begin{figure}[t]
\includegraphics[width=0.49\textwidth]{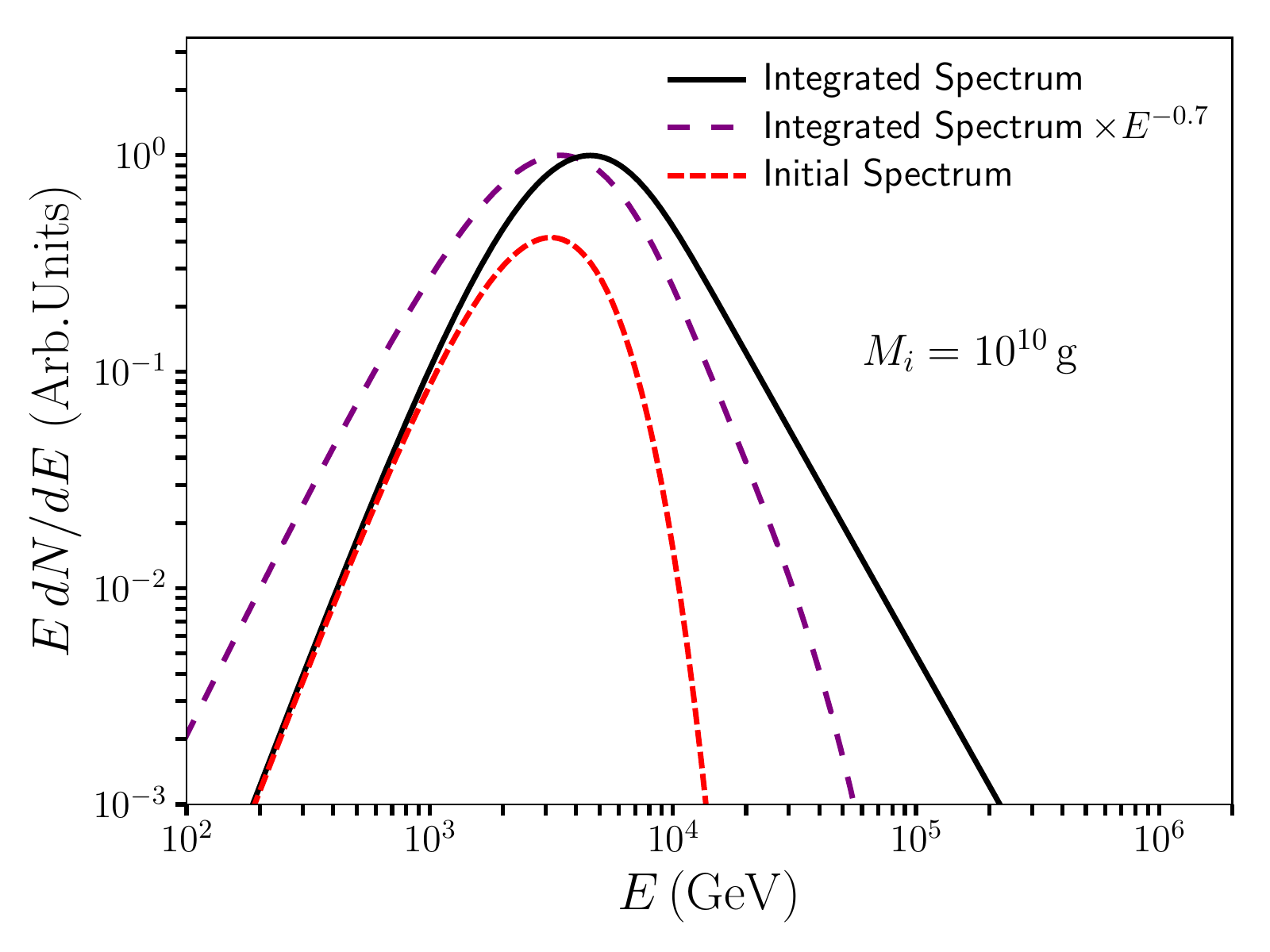} 
\includegraphics[width=0.49\textwidth]{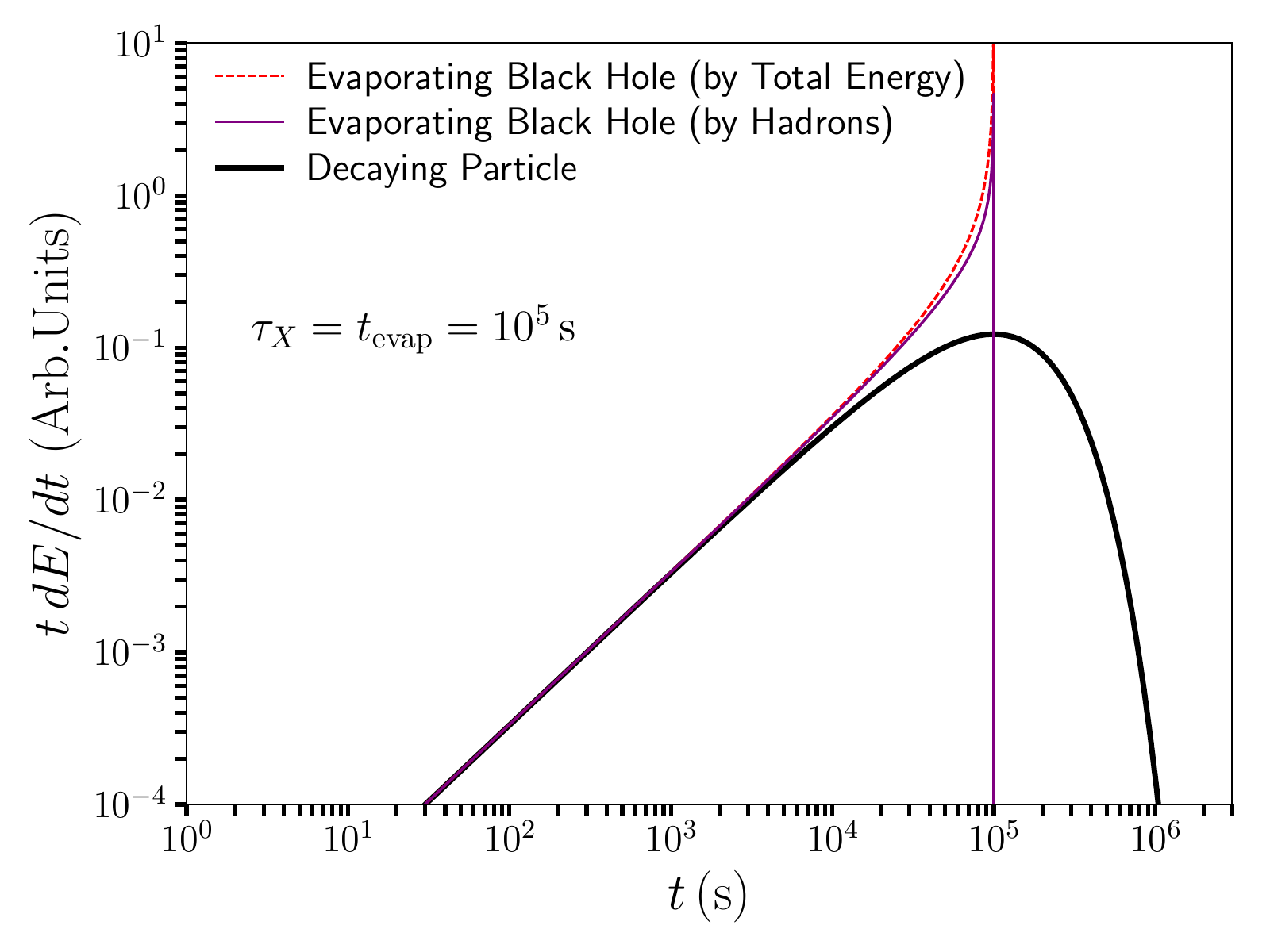} 
\caption{Left Frame: The spectrum of particles radiated from a black hole with an initial mass of $10^{10}$ grams. We show the initial spectrum (when $M=10^{10}$ g), the spectrum integrated over the lifetime of the black hole, and the integrated spectrum weighted by an additional factor of $E^{-0.7}$ (as appropriate for considering the production of hadrons). Right frame: The time profile for energy injection from particle decay or black hole evaporation, for the case of a lifetime or evaporation time of $10^5$ seconds. In the case of black hole evaporation, we show profiles corresponding to the total injected energy and to the number of injected hadrons. }
\label{timeprofile}
\end{figure}

A second way in which evaporating black holes impact BBN differently from long-lived particles follows from the fact that the temperature of a black hole (and thus the average energy of the injected particles) increases as a black hole loses mass. For example, when a particle decays into a pair of quarks, $X \rightarrow q\bar{q}$, those quarks each have an energy of $E_q = m_X/2$. Hawking radiation, in contrast, produces an approximately thermal spectrum of particles, with a temperature that steadily increases as the black hole radiates.\footnote{Throughout this study, we adopt a thermal (Fermi-Dirac or Bose-Einstein) distribution for the spectral shape of the products of Hawking evaporation. Although this is not precisely true~\cite{Page:1976df,1990PhRvD..41.3052M,1991PhRvD..44..376M}, it is an adequate approximation for the purposes of this study.}  Averaged over the course of a black hole's evaporation, the mean energy of a radiated quark (or other fermion) is given by:
\begin{equation}
\langle E_q \rangle = \frac{\int^0_{M_i} 3.15 \,T_{\rm BH}(M) \, \frac{dN}{dM}(M) \, dM}{\int^{0}_{M_i} \frac{dN}{dM}(M) \,dM} = 6.3 \, T_i,
\end{equation}
where $M_i$ ($T_i$) is the initial mass (temperature) of the black hole, $dN/dM \propto T^{-1}_{\rm BH}$ is the number of particles radiated per unit mass loss, and we have made use of the fact that the average energy of a relativistic fermion in a thermal distribution is approximately 3.15 times the temperature of that distribution.
Based on this result, we approximate the spectrum of the emission from an evaporating black hole with that from the (two-body) decays of a particle with a mass equal to $m_X \simeq 12.6 \, T_i$. In the left frame of Fig.~\ref{timeprofile}, we plot the spectrum of Hawking radiation injected from a black hole with an initial mass of $M_i = 10^{10}$ g, both at that moment, and as integrated over the course of its evaporation.

The prescription described in the previous paragraph is appropriate for cases in which the destruction of helium nuclei is dominated by photodissociation (the total quantity of injected electromagnetic energy sets the rate of photodissociation). During the hadrodissociation era ($T \gsim 0.4$ keV), however, the number of helium nuclei that are broken up instead scales with the number of energetic hadrons that are injected into the early universe. The average number of hadrons that are produced in the jet from a given quark is roughly proportional to $E_q^{0.3}$, and thus the average number of hadrons produced per unit energy is approximately proportional to $E_q^{-0.7}$~\cite{Kawasaki:2017bqm}. Due to this scaling, as a black hole loses mass and radiates increasingly high-energy particles, a smaller fraction of the radiated energy takes the form of hadrons. Over the course of a black hole's evaporation, the average hadron is produced by a quark of energy $\langle E_q \rangle \simeq 3.7 \, T_i$. Thus in the hadrodissociation era, the spectrum of the emission from an evaporating black hole can be approximately related to that from a (two-body) decaying particle with a mass of $m_X \simeq  7.4 \,T_i$. This is illustrated by the fact that purple dashed curve in the left frame of Fig.~\ref{timeprofile} peaks at a lower energy than the solid black curve, by a factor of $7.4/12.6 \approx 0.6$.

\begin{figure}[t]
\includegraphics[width=0.49\textwidth]{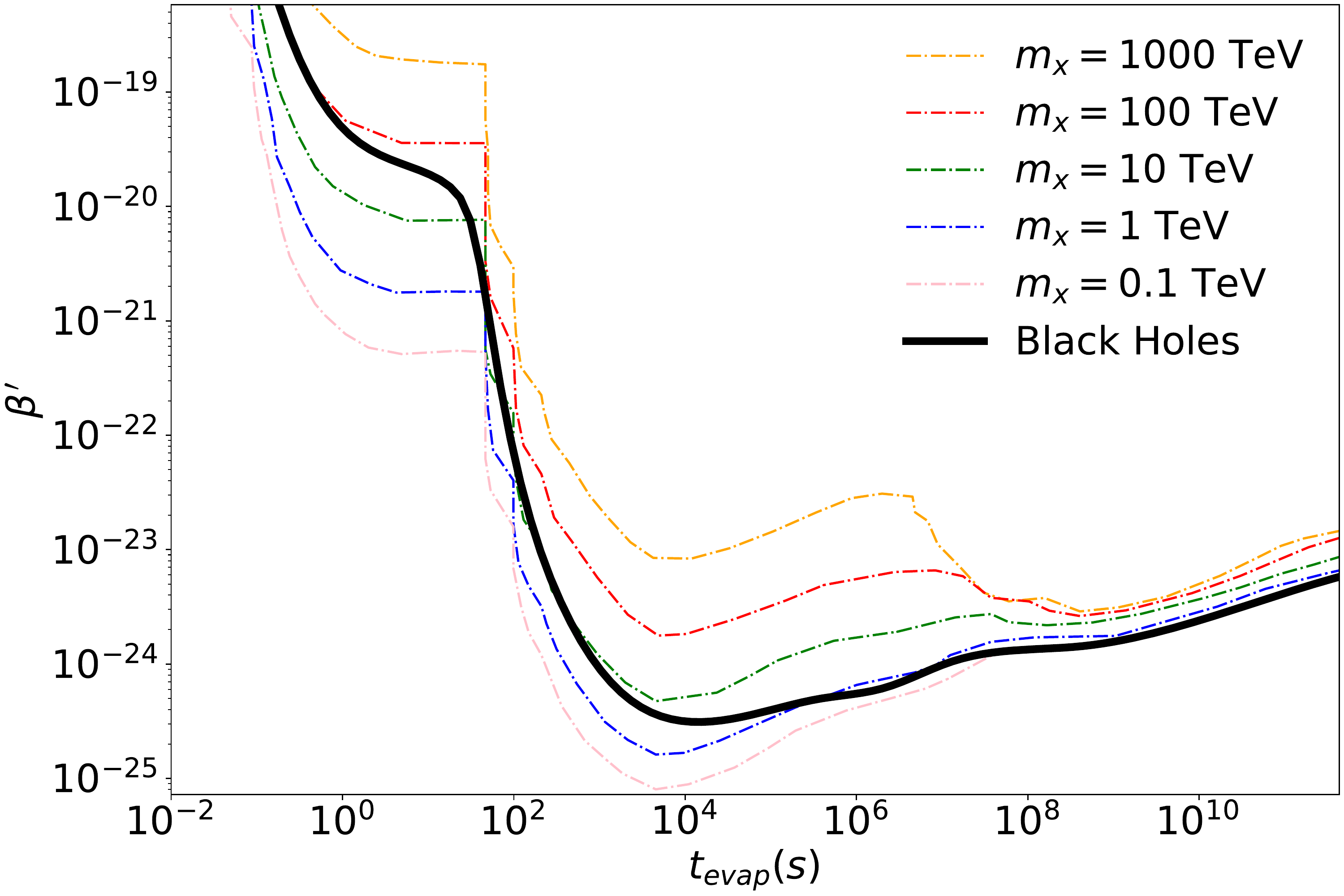} 
\includegraphics[width=0.49\textwidth]{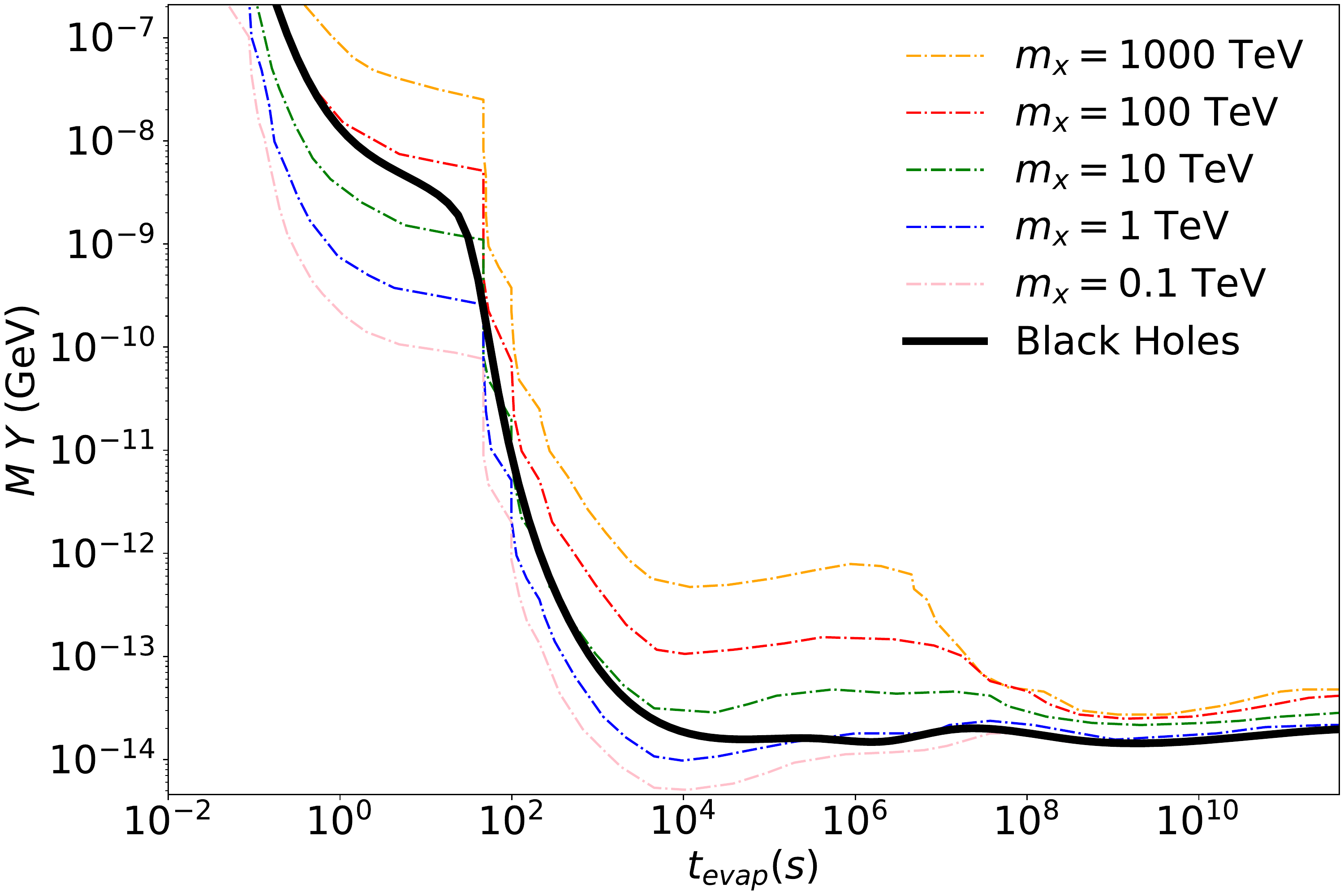} 
\caption{Constraints on long-lived particles from Ref.~\cite{Kawasaki:2017bqm}, for the case of $X \rightarrow q\bar{q}$, for several values of $m_X$.  These constraints are presented both in terms of $MY$, as used by Kawasaki {\it et al.}, (right frame) and in terms of $\beta'$, as used by Carr {\it et al}. (left frame). The solid black curve in each frame is our constraint on evaporating black holes, based on an interpolation between the long-lived particle constraints, following the relationship between $m_X$ and $T_i$ as described in the text. In this figure, we have assumed that the black holes evaporate only into Standard Model particles.}
\label{interpolation}
\end{figure}

A third way in which long-lived particles behave differently from evaporating black holes is in the rates at which they inject energetic particles into the early universe. Unlike a population of decaying particles, the evaporation rate of a black hole increases as it loses mass. In the right frame of Fig.~\ref{timeprofile}, we compare the time profiles for these emission mechanisms. Well before the particle's lifetime or black hole's evaporation time, the shape of these time profiles are nearly identical. During the final stages of evaporation and decay, however, they are quite different. In the photodissociation regime, in our translation of the constraints on decaying particles to the case of evaporating black holes, we shift the decaying particle's lifetime by a factor of 0.79 in order to match the time at which the mean unit of energy was injected into the early universe. In the hadrodissociation era, we instead adjust the lifetime such that the median hadron is injected at the same time. For decaying particles, this occurs
at a time, $t_{{\rm med}} = \ln 2 \times \tau_X = 0.69 \, \tau_X$, while for an evaporating black hole, $t_{{\rm med}} \simeq  0.71 \, t_{\rm evap}$. This case, therefore, requires only a small correction factor to translate between the two timescales.

In Fig.~\ref{interpolation}, we plot the constraints on long-lived particles from Ref.~\cite{Kawasaki:2017bqm}, for the case of $X \rightarrow q\bar{q}$, for several values of $m_X$.  These constraints are shown in terms of the quantities used by Kawasaki {\it et al.} (right frame), as well as those used by Carr {\it et al}. (left frame). Also shown as a solid black curve in each frame is our constraint on evaporating black holes, based on an interpolation between the long-lived particle constraints, following the relationship between $m_X$ and $T_i$ as described in the paragraphs above.

The procedure described in this section relies on the validity of two underlying assumptions. First, we have assumed that the overall shape of the spectrum of particles injected into the early universe does not strongly impact the resulting constraints, so long as the average energy is the same. Second, we have assumed that the time profile of the particle injection does not strongly impact the results, so long as the average particle is injected at the same time. We acknowledge that these assumption are not strictly true, and that these considerations could introduce a systematic error into the constraints that are presented here. In terms of the shape of the spectrum, considering the total integrated emission from a black hole (see Fig.~\ref{timeprofile}), approximately 32\% of the injected energy is in the form of particles that lie within only a factor of 2 in energy. Similarly, approximately 78\% of the injected energy is in particles that lie within an order of magnitude in energy. Combining this with the information shown in Fig.~\ref{interpolation}, we conclude that this could potentially introduce an error in our constraint that is as large as $\sim$\,30\% for $t_{\rm evap} \gsim 10^7 \, {\rm s}$, and up to a factor of $\sim$\,2 for shorter-lived black holes. On similar grounds, the more gradual time profile associated with the energy injection from the late stages of long-lived particle decay (see Fig.~\ref{timeprofile}) could potentially impact our constraints. For most values of $t_{\rm evap}$, this effect is quite small. For $t_{\rm evap} \approx 80-200\, {\rm s}$, however, the constraints change rapidly with $t_{\rm evap}$, allowing the resulting constraints to be impacted more significantly, potentially shifting the constraints on this part of parameter space by up to a factor of a few to the right (toward larger values of $t_{\rm evap}$).

\begin{figure}[t]
\includegraphics[width=0.49\textwidth]{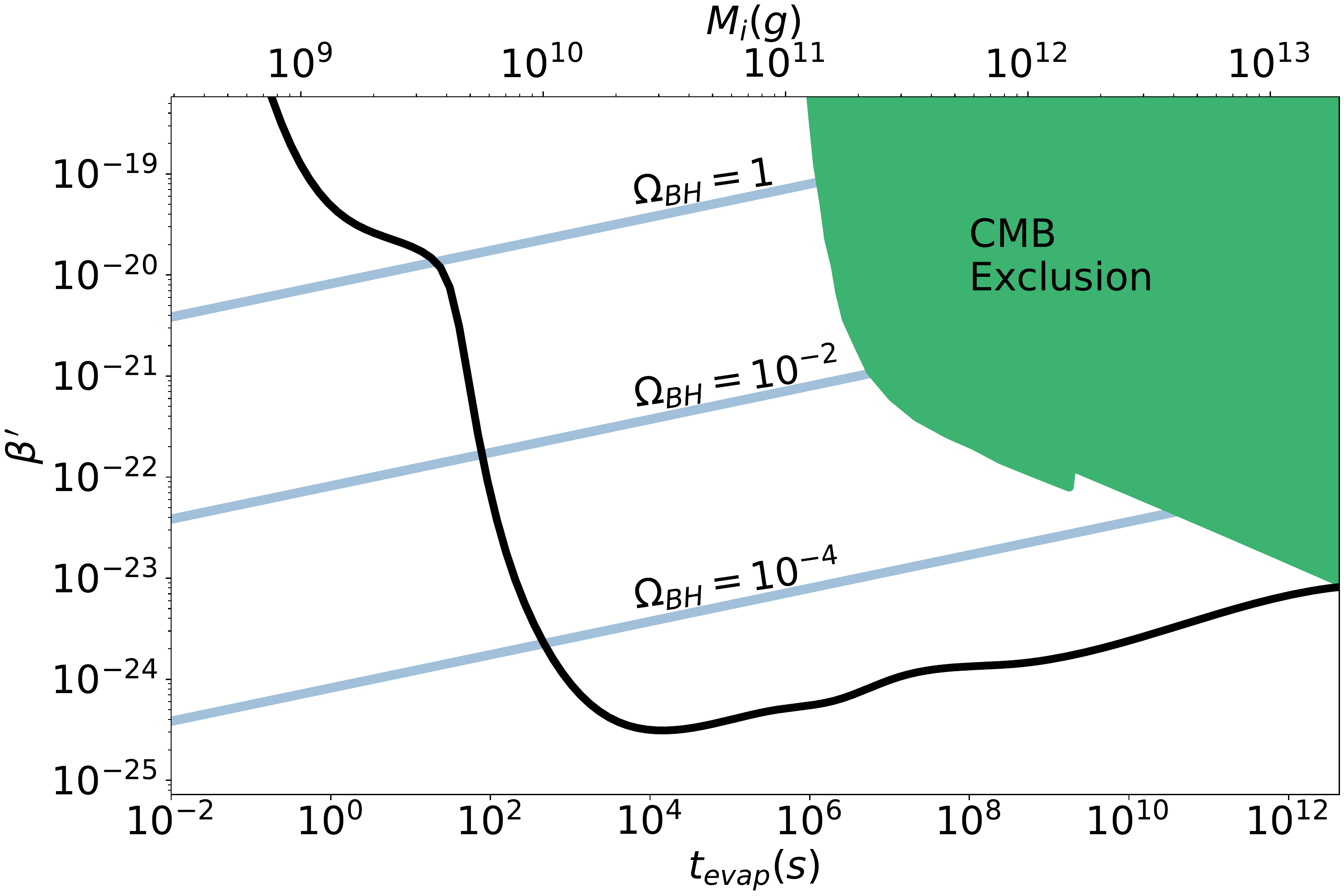} 
\includegraphics[width=0.49\textwidth]{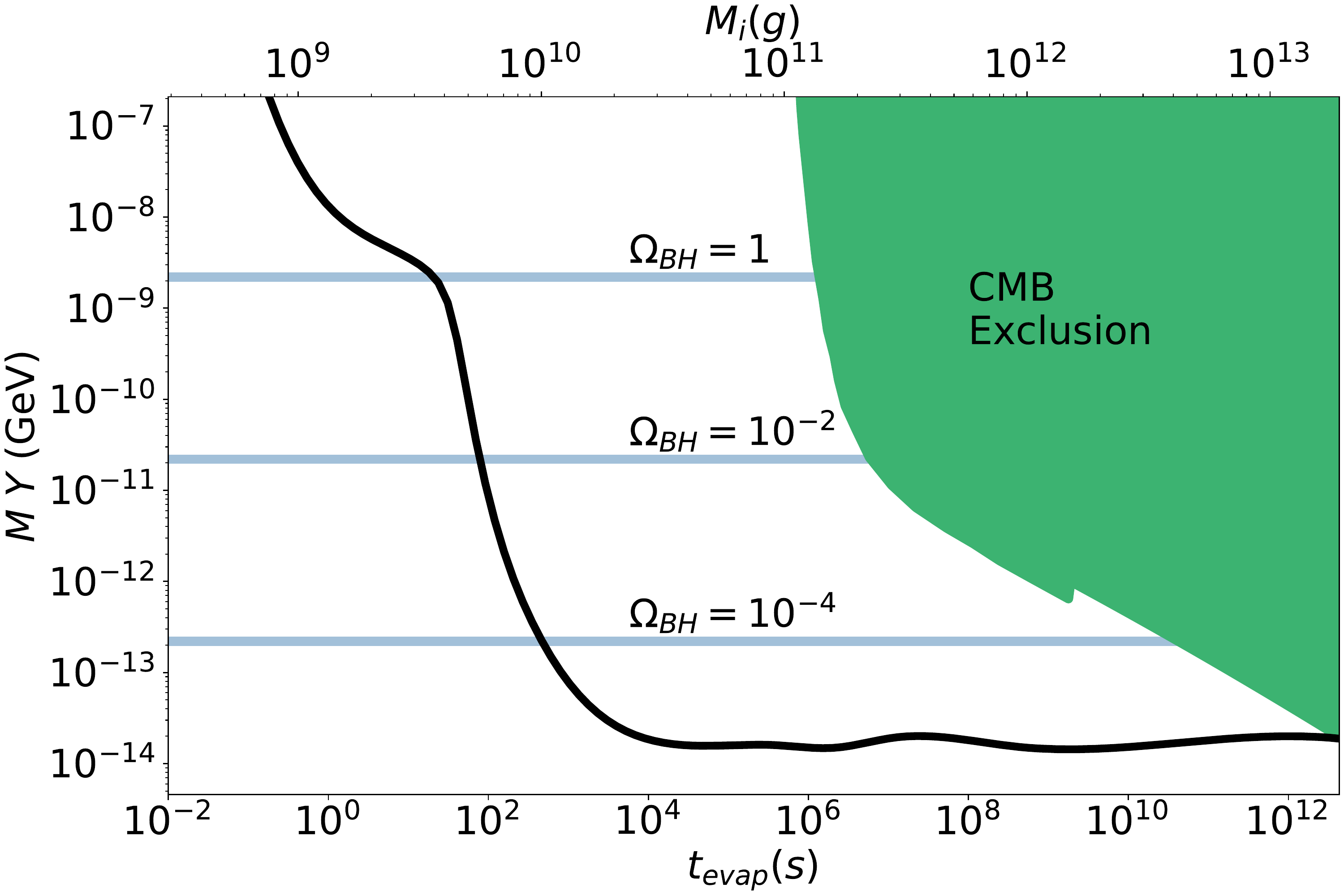} 
\caption{Constraints on primordial black holes, assuming that they evaporate entirely into Standard Model particles. Again, we have presented these constraints both in terms of $MY$, as used by Kawasaki {\it et al.}, (right frame) and in terms of $\beta'$, as used by Carr {\it et al}. (left frame).  For rapidly evaporating black holes ($t_{\rm evap} \lsim 80$ s), the constraints are dominated by the measured primordial helium fraction, $Y_p$, while for longer evaporation times the primordial deuterium abundance provides the most stringent constraint. In each frame, we also plot contours of constant $\Omega_{\rm BH}$, defined as the value of $\rho_{\rm BH}/\rho_{\rm crit}$ that would be the case today if the black holes had not evaporated. The green regions in the upper-right regions of each frame are excluded by measurements of the CMB (via spectral distortions).}
\label{mainSM}
\end{figure}

In Fig.~\ref{mainSM}, we plot our constraints on primordial black holes, assuming that they evaporate entirely into the particle content of the Standard Model (in other words, assuming that there is no particle content beyond the Standard Model). For rapidly evaporating black holes ($t_{\rm evap} \lsim 80$ s), these constraints are dominated by the measured primordial helium fraction, $Y_p$, while for longer evaporation times the primordial deuterium abundance provides the most stringent constraint. In each frame, we also plot contours of constant $\Omega_{\rm BH}$, defined as the value of $\rho_{\rm BH}/\rho_{\rm crit}$ that would be the case today if the black holes had not evaporated. In the upper right corner of each frame, we show constraints on evaporating black holes based on spectral distortions of the CMB, as derived in Ref.~\cite{Kawasaki:2017bqm} (see also Refs.~\cite{Chluba:2011hw,Chluba:2013wsa,Poulter:2019ooo, Acharya:2019xla,Stocker:2018avm,Stocker:2018avm,Clark:2016nst}). We note that constraints derived from CMB spectral distortions due to evaporating primordial black holes have recently been revisited in somewhat more detail~\cite{Acharya:2019xla, Chluba:2020oip}, resulting in somewhat weaker bounds. Future measurements by PIXIE are expected to improve upon these constraints by a factor of $\sim 10^3$ or more~\cite{Chluba:2013pya, Chluba:2020oip}. Primordial black holes with somewhat higher masses, which evaporate slightly after the formation of the CMB, may also be constrained by considering their effects on the CMB power spectrum and the optical depth to reionization~\cite{Poulin:2016anj, Stocker:2018avm}.

\section{Constraints on Black Holes in Scenarios Beyond the Standard Model}
\label{BSM}

Unlike particle decays (and most other processes in nature), Hawking evaporation is an entirely gravitational phenomenon, and thus produces all particle species with masses below $\sim T_{\rm BH}$, regardless of their charges or couplings. As a result, the rate at which Hawking evaporation occurs, and the varieties of particles that are produced through this process, depend on the complete spectrum of particles that exist, including all such species beyond the limits of the Standard Model (for previous literature that has explored such possibilities, see Refs.~\cite{Hooper:2019gtx,Hooper:2020evu,Lennon:2017tqq,Lunardini:2019zob,Masina:2020xhk,Morrison:2018xla,Allahverdi:2017sks,Hamada:2016jnq,Fujita:2014hha,Hook:2014mla}).

The existence of physics beyond the Standard Model can impact the constraints presented in this paper in three ways. First, additional particle species have the effect of increasing the rate at which black holes evaporate, shifting (and typically weakening) the resulting limits. Second, any particle species without appreciable couplings to the Standard Model will only impact the light element abundances through their influence on the expansion history of the universe. In scenarios that include large numbers of decoupled degrees-of-freedom, the fraction of a black hole's mass that goes into particles that can break up helium and produce deuterium can be significantly reduced, while instead producing substantial abundances of dark matter and/or dark radiation~\cite{Hooper:2019gtx,Hooper:2020evu,Lennon:2017tqq,Masina:2020xhk,Morrison:2018xla,Allahverdi:2017sks,Fujita:2014hha,Baldes:2020nuv}. Third, the presence of black holes and their decoupled evaporation products can impact the expansion history of the early universe, altering the light element abundances that emerge from this era without directly disrupting any nuclei. In the remainder of this section, we will explore several classes of scenarios beyond the Standard Model and discuss their impact on the resulting constraints on primordial black holes.

A wide range of well-motivated scenarios have been proposed in which the degrees-of-freedom associated with the Standard Model constitute only a small fraction of the particle spectrum of the universe. In particular, self-consistent string compactifications have been shown to generically predict the existence of large numbers of feebly interacting states, including gauge bosons, axion-like particles, and other forms of exotic matter~\cite{Cvetic:2011iq,Halverson:2018vbo,Halverson:2016nfq,Svrcek:2006hf,Svrcek:2006yi,Arvanitaki:2009fg,Fox:2004kb,Braun:2005bw,Cleaver:1998gc,Coriano:2007ba,Giedt:2000bi,Lebedev:2007hv,Anastasopoulos:2006cz,Cvetic:2001nr}. Frameworks featuring extremely large numbers of massive degrees-of-freedom have also been considered within the context of possible solutions to the electroweak hierarchy problem~\cite{Dvali:2007hz,Dvali:2009ne,Arkani-Hamed:2016rle}.

Given the gravitational nature of Hawking radiation, black holes are expected to radiate all particle species with masses below $\sim T_{\rm BH}$, regardless of their charges or couplings. Thus in scenarios with expansive particle content, black holes could potentially radiate mostly or almost entirely to hidden sector states, which could act as a combination of dark matter and dark radiation~\cite{Hooper:2019gtx,Hooper:2020evu,Lennon:2017tqq,Masina:2020xhk,Morrison:2018xla,Allahverdi:2017sks,Fujita:2014hha,Baldes:2020nuv}. If feebly interacting, such particles would not directly disrupt nuclei during or after BBN, but could still impact the resulting light element abundances through their impact on the universe's expansion rate during this era.

To constrain a black hole population that evaporates dominantly to hidden sector particles, we calculate the combined energy density of the black holes and their evaporation products as a function of redshift, and then use the publicly available program \alterbbn~\cite{Arbey:2011nf,Arbey:2018zfh} to calculate the resulting light element abundances. In doing so, we follow the procedure described in Ref.~\cite{Berlin:2019pbq}, and use the deuterium-burning rates from Ref.~\cite{Coc:2015bhi} and other reaction rates from Refs.~\cite{Ando:2005cz,Cyburt:2004cq,Xu:2013fha}. These rates correspond to systematic uncertainties of 1.9\% on $(\rm{D}/\rm{H})_p$ and 0.13\% on $Y_p$, approximately independent of the time evolution of the energy injection~\cite{Berlin:2019pbq}. 

To calculate the evolution of the energy densities in black holes and their evaporation products, we solve the following system of differential equations:
\begin{eqnarray}
\label{rhoM2}
\frac{d\rho_{\rm BH}}{dt}&=& -3\rho_{\rm BH}H+\rho_{\rm BH}\frac{dM}{dt}\frac{1}{M},  \\
\frac{d\rho_{\rm SM}}{dt}&=& -3(w_{\rm SM}+1)\rho_{\rm SM}H-\rho_{\rm BH}\frac{dM}{dt}\frac{(1 - f_d)}{M},  \nonumber \\
\frac{d\rho_{d}}{dt}&=& -3(w_d+1)\rho_{d}H-\rho_{BH}\frac{dM}{dt}\frac{f_d}{M},  \nonumber 
\end{eqnarray}
where $\rho_{\rm BH}$, $\rho_{\rm SM}$, $\rho_{d}$ are the energy densities in black holes, Standard Model fields, and dark matter plus dark radiation, respectively, $H^2 = 8\pi G (\rho_{\rm BH}+\rho_{\rm SM}+\rho_{d})/3$ is the rate of Hubble expansion, and $dM/dt$ is the black hole evaporation rate (see Eq.~\ref{loss}). The quantities $w_{\rm SM}$ and $w_{d}$ represent the equation-of-state of the Standard Model and hidden sector particles, with values of 0 and 1/3 corresponding to pure matter and radiation, respectively. Lastly, $f_d$ is the fraction of Hawking radiation that proceeds to hidden sector particles, following from the Standard Model and hidden sector contributions to $g_{\star, H}$. The temperature dependence of $w_{\rm SM}$ is directly related to the values of $g_{\star}$ and $g_{\star, S}$ (for more details, see Refs.~\cite{Kolb:1990vq,Hooper:2019gtx}). 

\subsection{Light Hidden Sectors}
\label{hiddenlight}

\begin{figure}[t]
\includegraphics[width=0.49\textwidth]{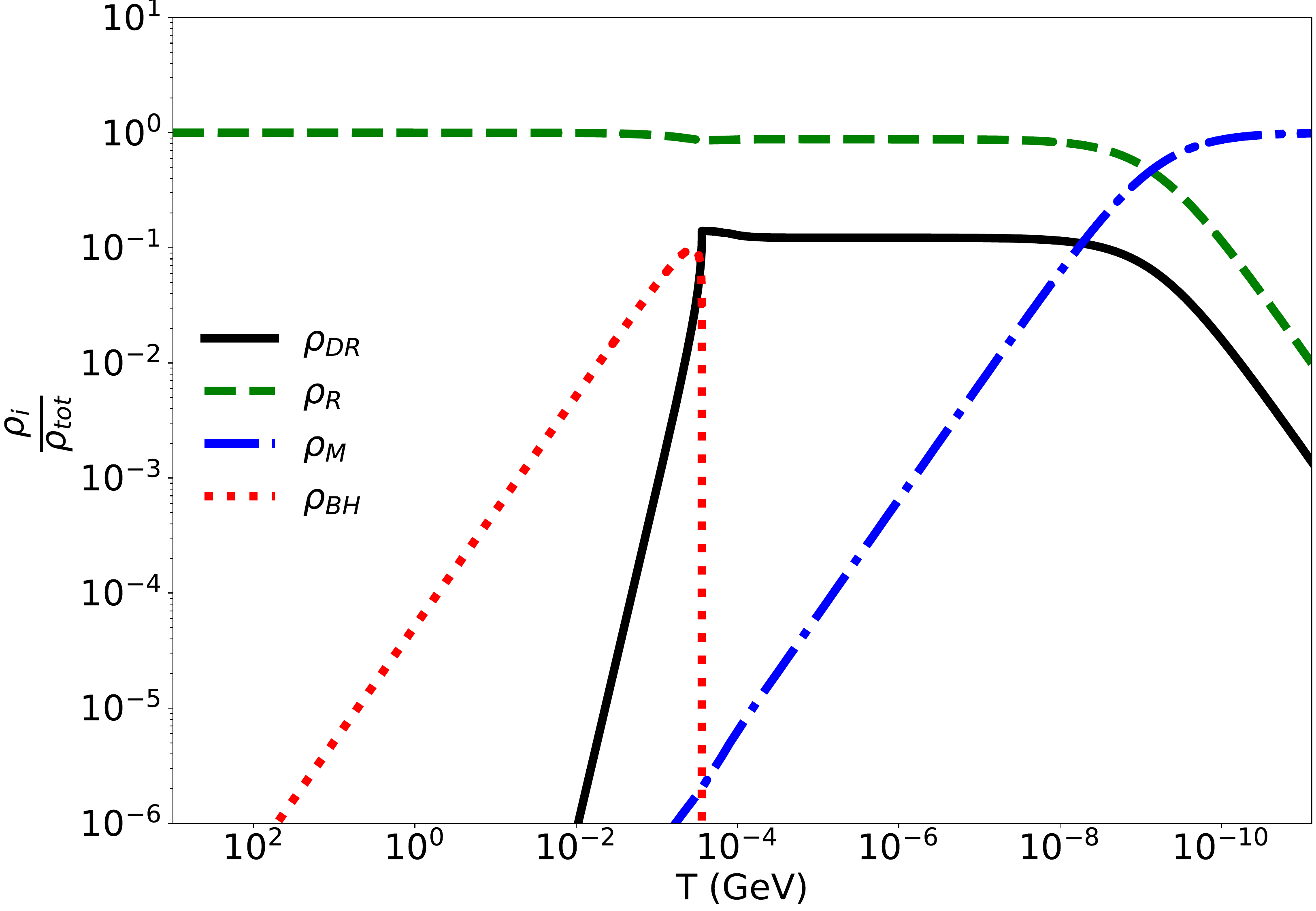} 
\caption{The evolution of the energy densities in black holes, Standard Model radiation, matter (including both baryonic and dark matter), and dark radiation, in a scenario in which the black holes evaporate almost entirely to dark radiation (corresponding to $w_d=1/3$ and $f_d \simeq 1$). We have adopted an evaporation time of $t_{\rm evap}=10 \, {\rm s}$ and an initial black hole abundance corresponding to $\Omega_{\rm BH} = 2.6\times 10^4$ (defined as the value that would be the case today if the black holes had not evaporated). In this scenario, the final ($t \gg t_{\rm evap}$) energy density of dark radiation corresponds to $\Delta N_{\rm eff} = 1.0$.}
\label{evol1} 
\end{figure}

We begin by considering a class of scenarios in which the black holes evaporate almost entirely to light, hidden sector states (corresponding to $w_d=1/3$ and $f_d \simeq 1$, which implies $g_{\star, H} \gg 10^2$). In Fig.~\ref{evol1}, we plot the evolution of the energy densities in black holes, Standard Model radiation, matter (including both baryonic and dark matter), and dark radiation, for the case of $t_{\rm evap}=10 \, {\rm s}$ and an initial black hole abundance corresponding to $\Omega_{\rm BH} = 2.6\times 10^4$ (defined as the value that would be the case today if the black holes had not evaporated). To relate the value of $\Omega_{\rm BH}$ to that of $\beta'$ or $MY$, see Eqs.~\ref{carrkawasaki} and~\ref{omegabh}. In this scenario, the black hole population evaporates almost entirely into dark radiation at $t \sim t_{\rm evap} = 10 \, {\s}$. The ultimate energy density of this dark radiation, which we determine by solving Eq.~\ref{rhoM2}, can be written in terms of its contribution to the effective number of neutrino species, $\Delta N_{\rm eff}$ (as evaluated at $t \gg t_{\rm evap}$):
\begin{equation}
    \label{Neff}
    \Delta N_{\rm eff} = \frac{\rho_{\rm DR}}{\rho_{\rm R}}\left[N_{\nu}+\frac{8}{7}\left( \frac{11}{4}\right)^{4/3}\right],
\end{equation}
where $N_\nu = 3.046$, $\rho_{\rm DR}$ is the energy density of dark radiation, and $\rho_{\rm R}$ is the energy density in photons and neutrinos. In the scenario shown in Fig.~\ref{evol1}, the energy density of dark radiation corresponds to a value of $\Delta N_{\rm eff} = 1.0$. In more generality, the contribution to $\Delta N_{\rm eff}$ from black hole evaporation (in the $w_d=1/3$ and $f_d \simeq 1$ limit) is given by:
\begin{equation}
    \label{Neff}
    \Delta N_{\rm eff} \approx 1.0 \times \bigg(\frac{\Omega_{\rm BH}}{2.6 \times 10^4}\bigg) \, \bigg(\frac{t_{\rm evap}}{10 \, {\rm s}}\bigg)^{1/2} \, \bigg(\frac{10}{g_{\star}(T_{\rm evap})}\bigg) \, \bigg(\frac{g_{\star, S} (T_{\rm evap})}{10}\bigg)^{4/3}, 
\end{equation}
where $g_{\star}(T_{\rm evap})$ and $g_{\star, S}(T_{\rm evap})$ are the effective numbers of relativistic degrees-of-freedom and relativistic degrees-of-freedom in entropy, respectively, each evaluated at the temperature at $t_{\rm evap}$.


An observant reader may notice a small bump-like feature in the dark radiation curve near $T\sim 10^{-4} \, {\rm GeV}$ in Fig.~\ref{evol1}. This feature is due to an entropy dump that occurs among the Standard Model particles in the thermal bath. Whereas the dark radiation energy density simply evolves with four powers of the scale factor, $\rho_{\rm DR} \propto a^{-4}$, the Standard Model ``radiation'' includes particles with non-negligible mass, and thus the energy density of this component evolves as $\rho_{\rm R} \propto a^{-4} \,g_{\star}/g_{\star, S}^{4/3}$, where $g_{\star}$ is effective number of relativistic degrees-of-freedom and $g_{\star, S}$ is the effective number of relativistic degrees-of-freedom in entropy. As the temperature decreases, $g_{\star}/g_{\star, S}^{4/3}$ increases, reducing the ratio of dark radiation to Standard Model radiation (for a more detailed discussion, see Sec.~III of Ref.~\cite{Hooper:2019gtx}).

\begin{figure}[t]
\includegraphics[width=0.49\textwidth]{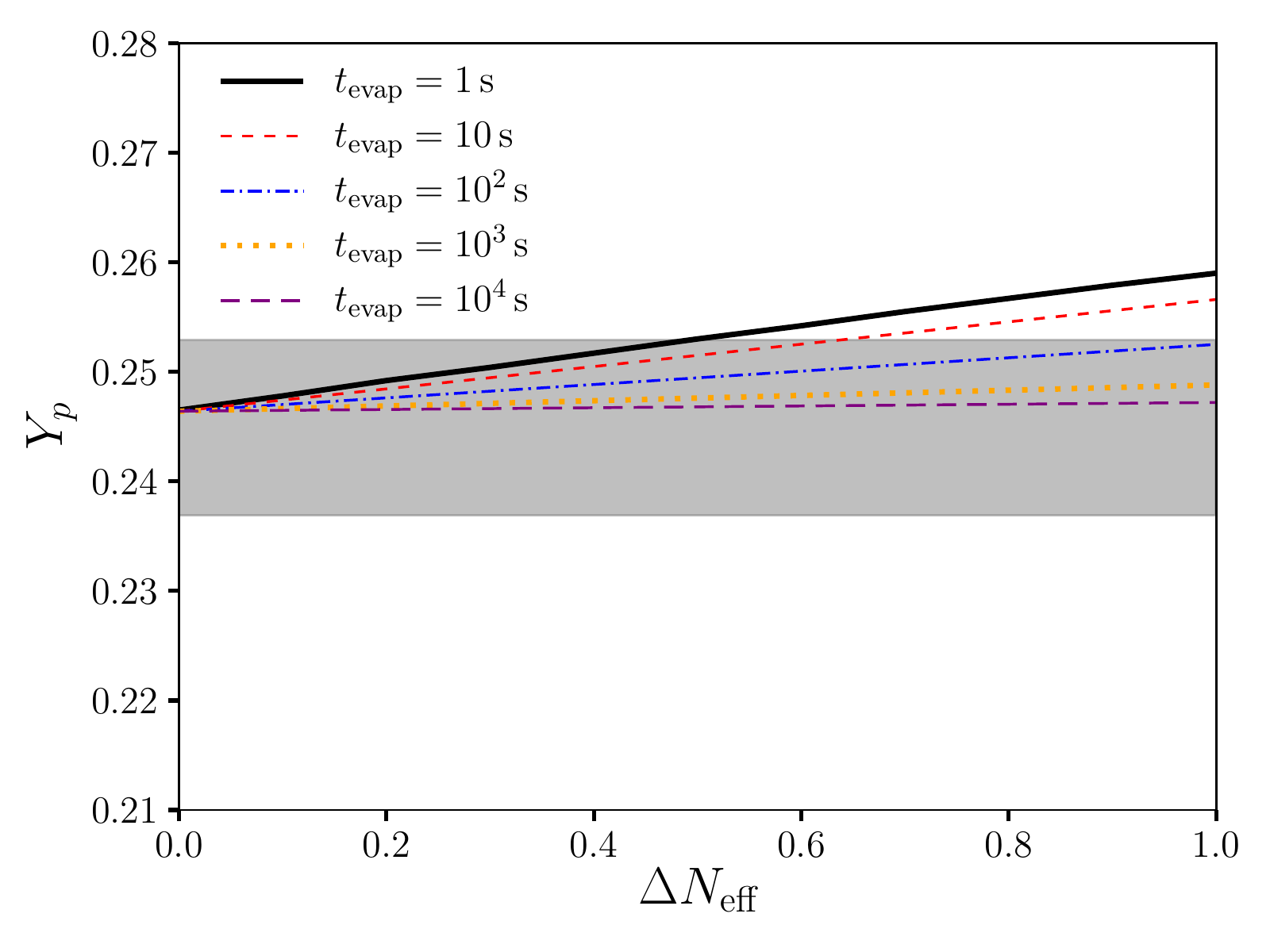} 
\includegraphics[width=0.49\textwidth]{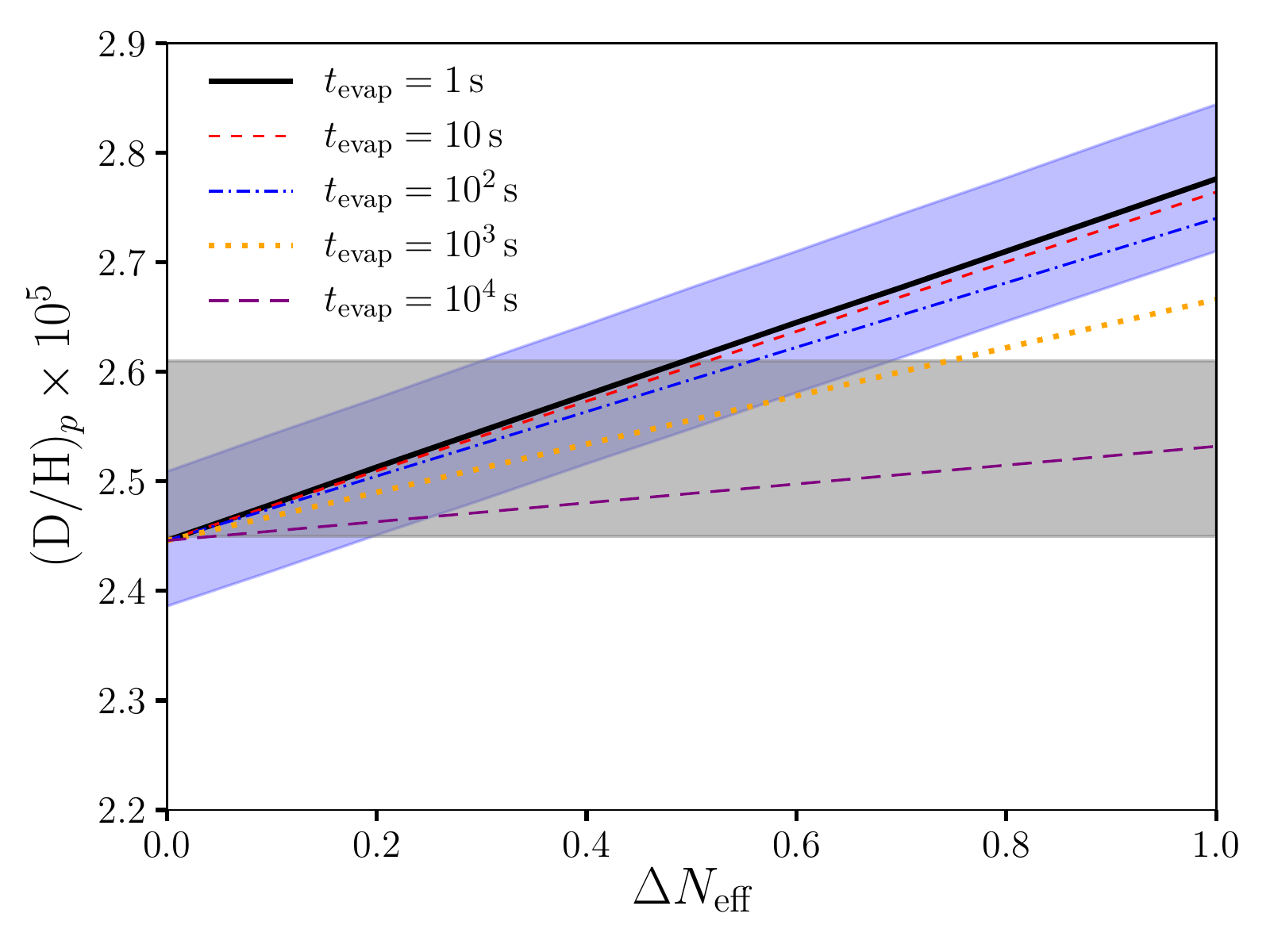} 
\caption{The impact on the primordial helium (left) and deuterium (right) abundances of black holes that evaporate overwhelmingly to dark radiation ($f_d \simeq 1$, $w_d \simeq 1/3$). These results are given in terms of the final ($t \gg t_{\rm evap}$) energy density of dark radiation, in terms of $\Delta N_{\rm eff}$. The grey bands represent the measured values (at $2\sigma$), while the blue band in the right frame denotes the systematic uncertainty associated with the nuclear reaction rates (as described in Sec.~\ref{BSM}). Note that this systematic uncertainty applies to all of the curves shown in the right frame (but for clarity is plotted only for the $t_{\rm evap} = 1 \, {\rm s}$ case). For relatively short-lived black holes ($t_{\rm evap} \lsim 10^2 \, {\rm s}$), the measured helium and deuterium abundances rule out scenarios in which this component of dark radiation contributes more than $\Delta N_{\rm eff} \gsim 0.4-0.6$.}
\label{darkradiation}
\end{figure}

In this class of scenarios, the black holes and their dark radiation evaporation products impact the light element abundances almost entirely through their influence on the expansion history of the early universe. In Fig.~\ref{darkradiation}, we illustrate the impact of such black holes on the primordial helium and deuterium abundances, as a function of the final ($t \gg t_{\rm evap}$) energy density in dark radiation, written in terms of $\Delta N_{\rm eff}$. The resulting light element abundances are a function of $t_{\rm evap}$, and those cases with $t_{\rm evap} \lsim 1 \, {\rm s}$ asymptote to the case of  a constant $\Delta N_{\rm eff}$, while longer lifetimes impact the expansion history primarily at somewhat later times. For relatively short-lived black holes ($t_{\rm evap} \lsim 10^2 \, {\rm s}$), the measured helium and deuterium abundances rule out scenarios in which the dark radiation contributes more than $\Delta N_{\rm eff} \gsim 0.4-0.6$ (at the 95\% confidence level), similar to the constraints derived from measurements of the CMB~\cite{Aghanim:2018eyx}. For longer-lived black holes, the constraints on the resulting contribution to $\Delta N_{\rm eff}$ are weaker (although the constraints derived from the CMB are approximately equally stringent for evaporation times up to $t_{\rm evap} \sim 10^{12} \, {\rm s}$~\cite{Bringmann:2018jpr}). Written in terms of $\beta'$, the measured helium and deuterium abundances provide a constraint of $\beta' \lsim (0.8-6.7)\times 10^{-16} \times (g_{\star, H}/10^5)^{1/6}$ across this entire range of evaporation times considered here. Note that these constraints are much less stringent than those presented in Fig.~\ref{mainSM} (for the case of Standard Model particle content). From this comparison, we conclude that the constraints based on dark radiation Hawking evaporation products will be more stringent than those resulting from proton-neutron conversion or helium disruption only if $t_{\rm evap} \lsim 0.1 \, {\rm s}$, or if $t_{\rm evap} \lsim 10^2 \, \rm{s}$ and the particle content of the dark sector is very large, corresponding to $g_{\star, H} \gsim 10^5$ or greater.

\subsection{Heavy Hidden Sectors}
\label{hiddenheavy}

In this subsection, we will continue to study models which feature a large number of hidden sector states, focusing on Hawking radiation in the form of hidden sector particles with non-negligible masses (which thus contribute to the universe's dark matter abundance). To this end, we follow the same procedure described earlier in this section, but introduce $T_{\rm BH}$-dependent contributions to $g_{\star, H}$, accounting for the inability of a black hole to radiate particles that are much more massive than its temperature.

\begin{figure}[t]
\includegraphics[width=0.49\textwidth]{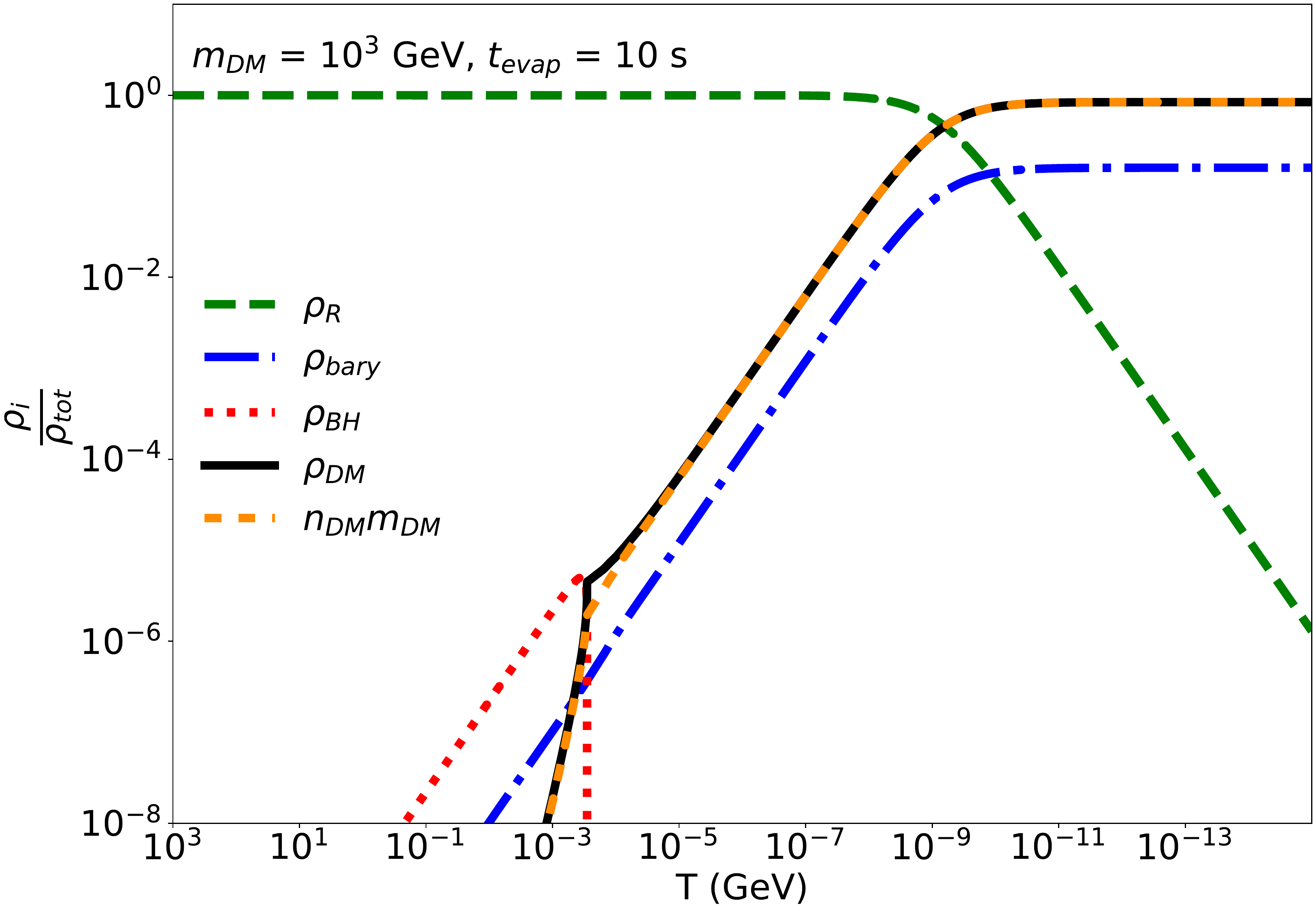} 
\includegraphics[width=0.49\textwidth]{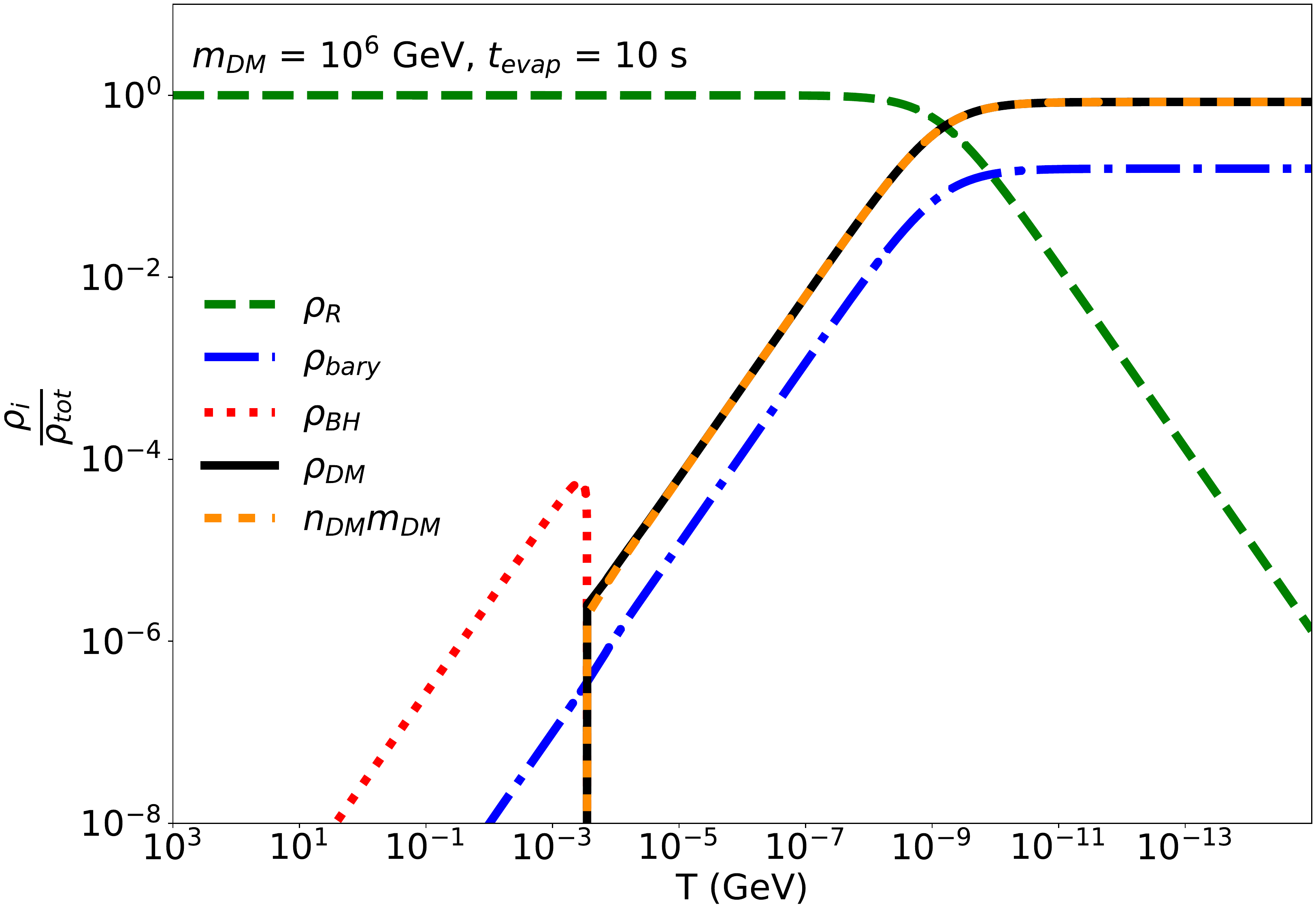} \\
\vspace{0.5cm}
\includegraphics[width=0.49\textwidth]{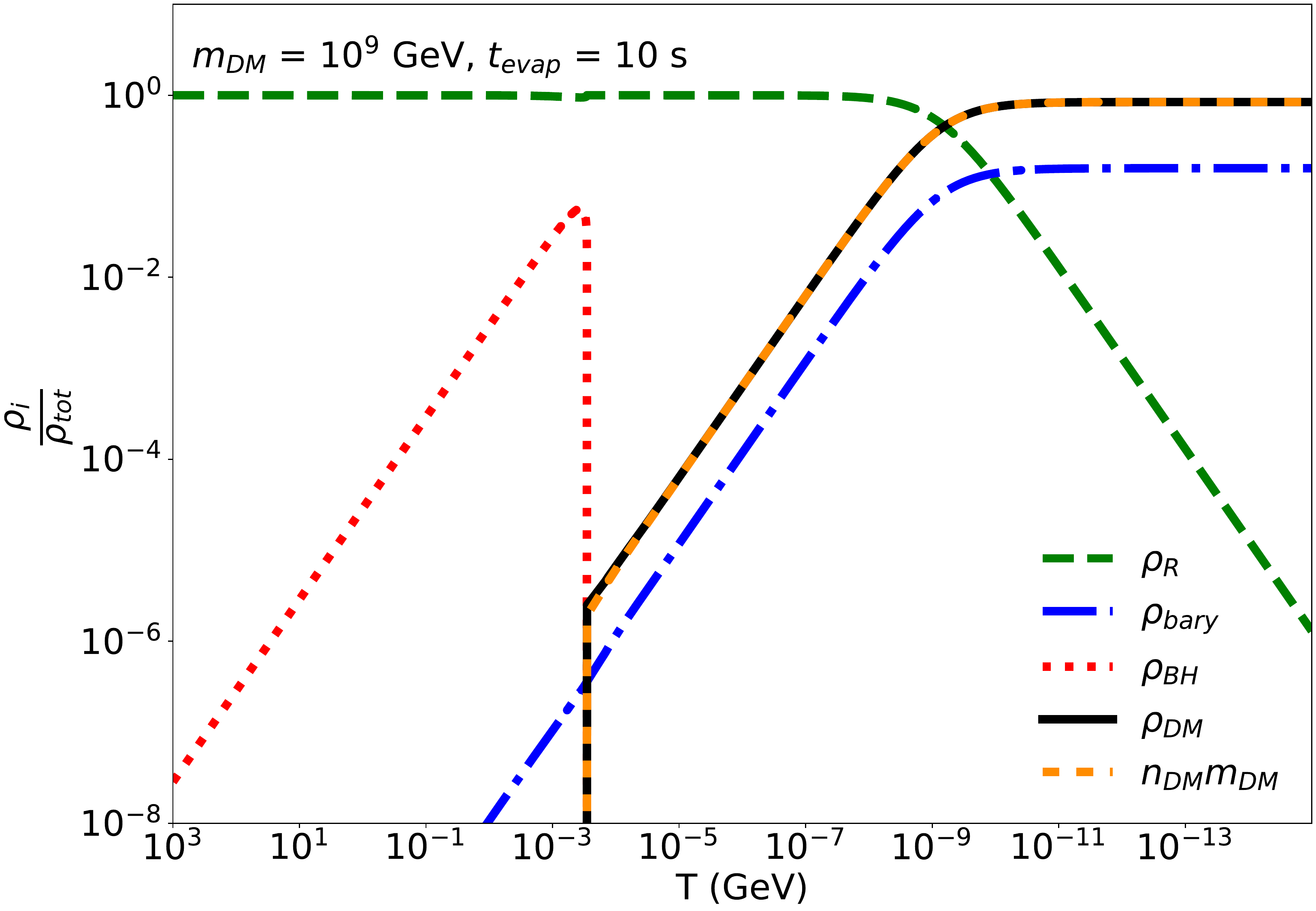}
\includegraphics[width=0.49\textwidth]{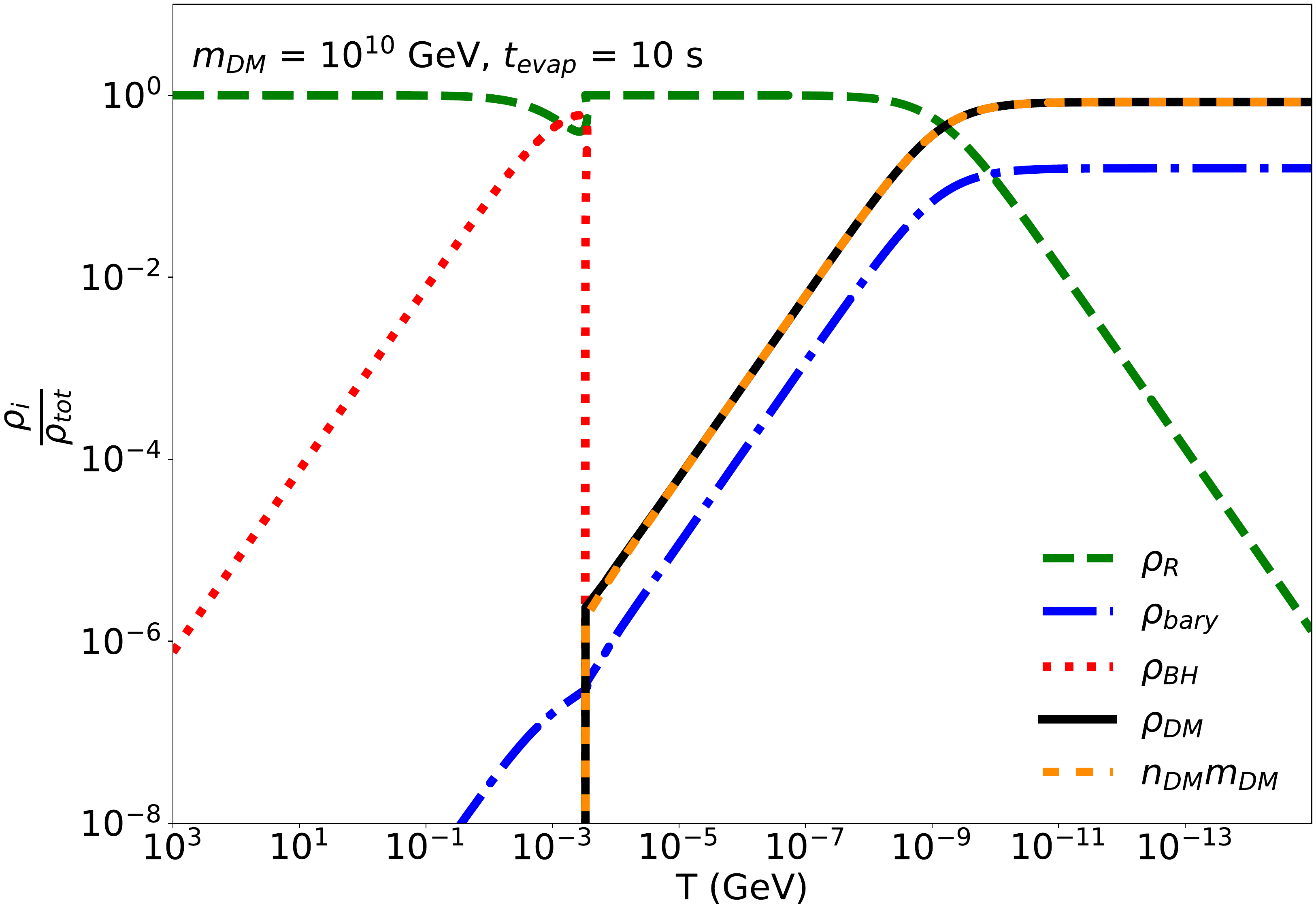} 
\caption{The evolution of the energy densities in Standard Model radiation, baryons, black holes, and dark matter, in a scenario in which the black holes evaporate with a lifetime of 10 seconds almost entirely to dark matter particles (corresponding to $g_{\star, H}=10^6$ for $T_{\rm BH} \gg m_{\rm DM}$). In each frame, the initial black hole abundance was chosen such that the Hawking radiation produces the entirety of the measured dark matter density. This corresponds to $\Omega_{\rm BH} = 6.8$ (upper left), 88 (upper right), $8.6\times 10^4$ (lower left) and $8.6 \times 10^5$ (lower right). As we have throughout this paper, we define $\Omega_{\rm BH}$ as the value that would be the case today if the black holes had not evaporated.}
\label{evolDM}
\end{figure}

\begin{figure}[t]
\includegraphics[width=0.49\textwidth]{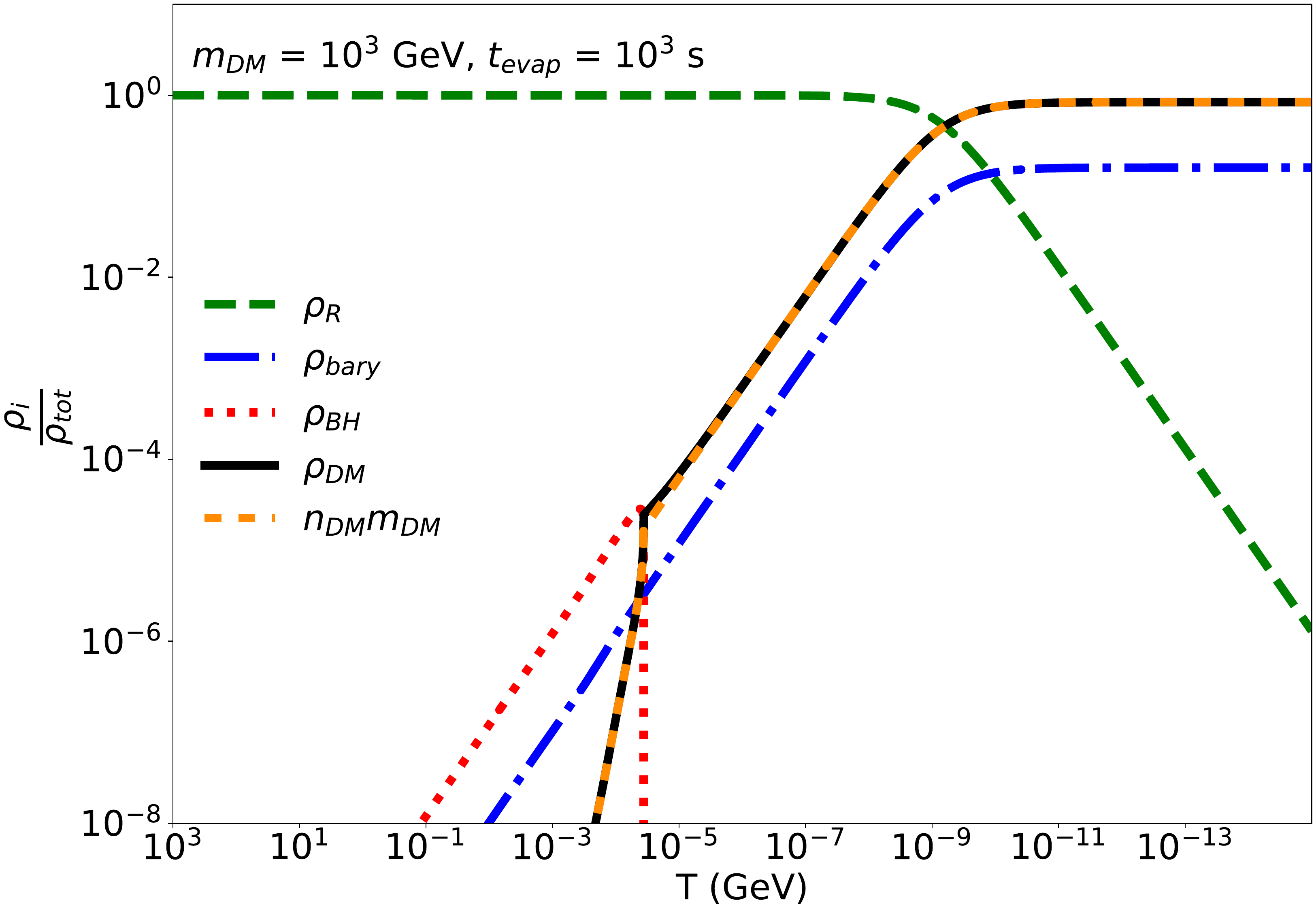} 
\includegraphics[width=0.49\textwidth]{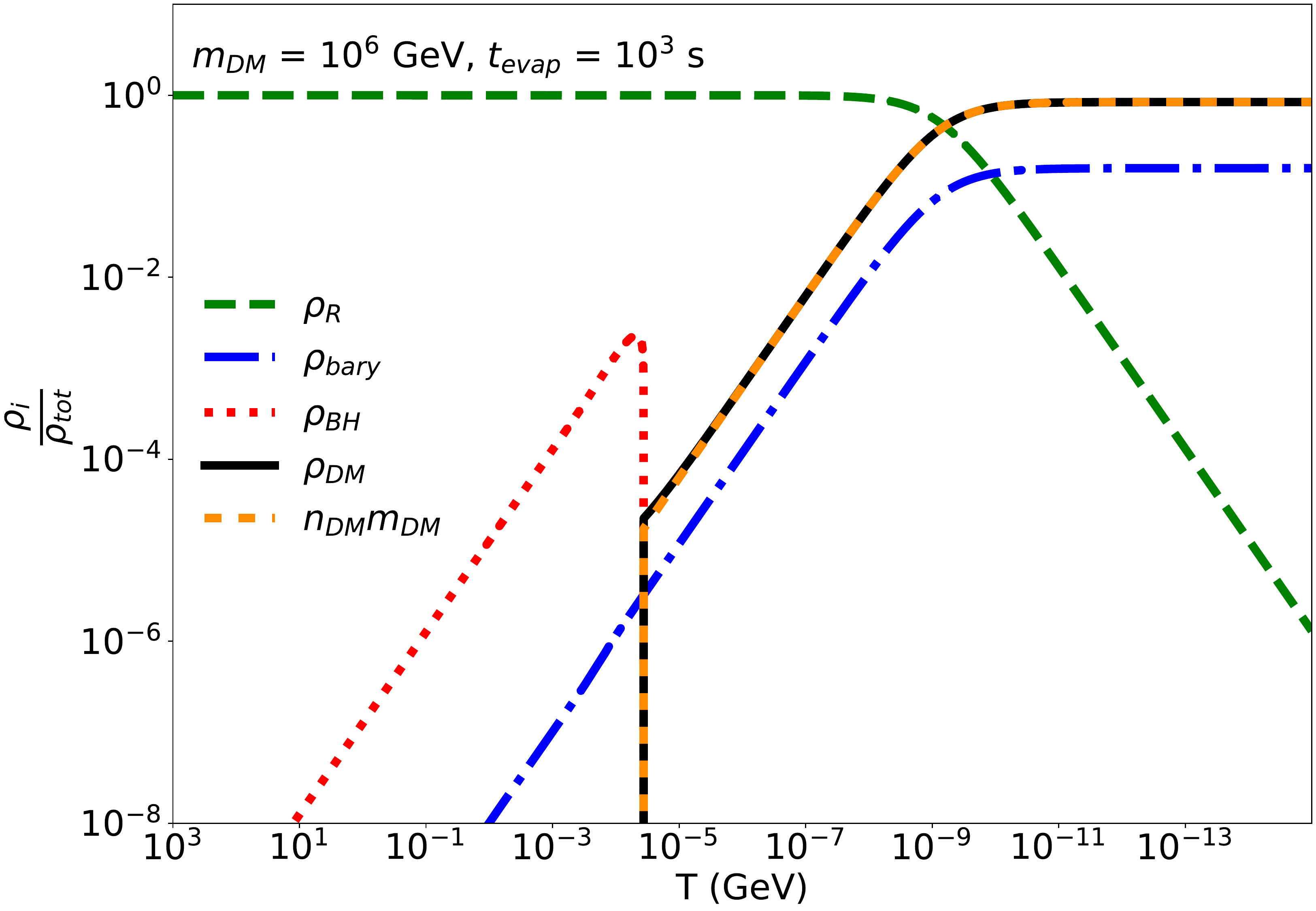} \\
\vspace{0.5cm}
\includegraphics[width=0.49\textwidth]{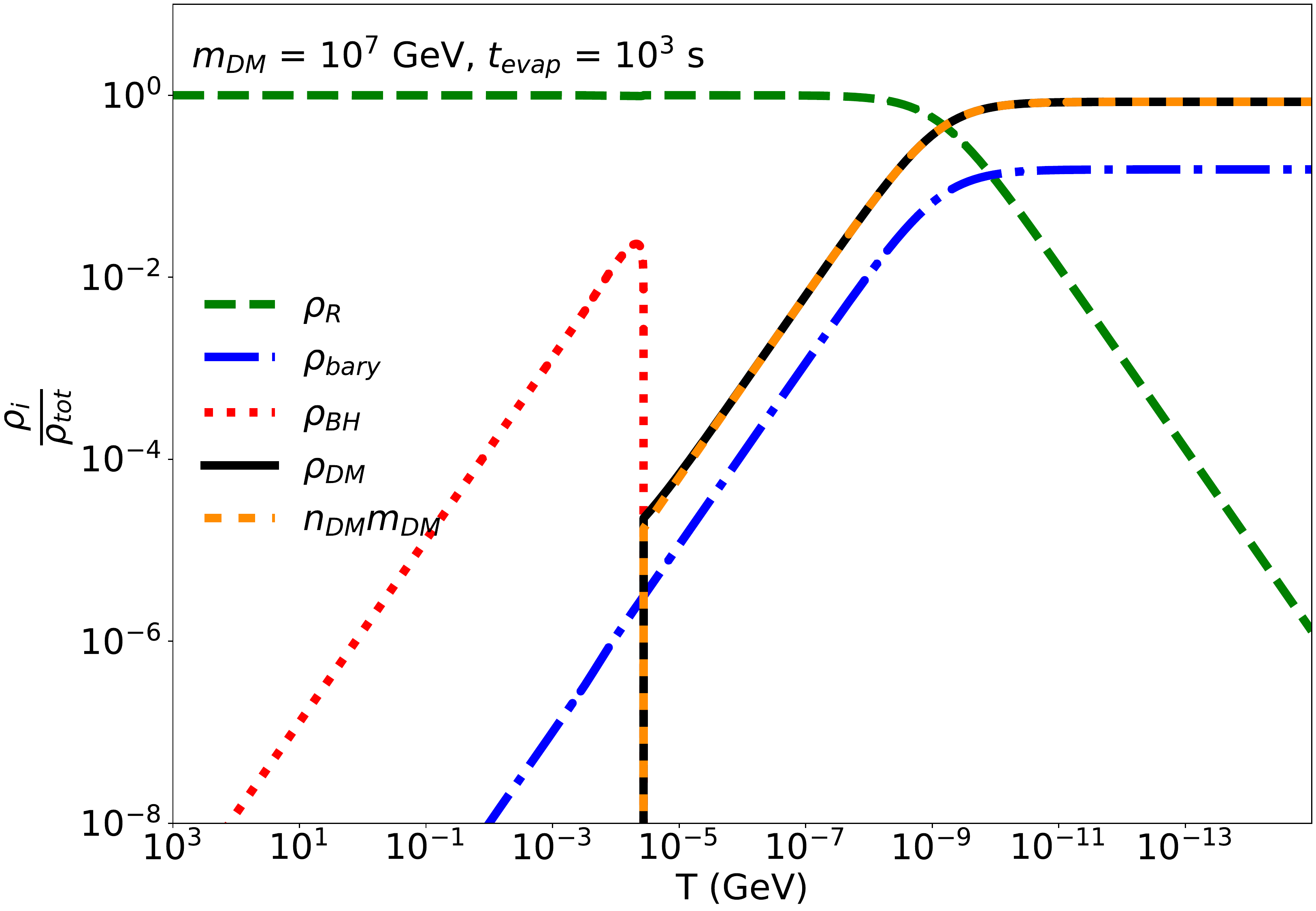}
\includegraphics[width=0.49\textwidth]{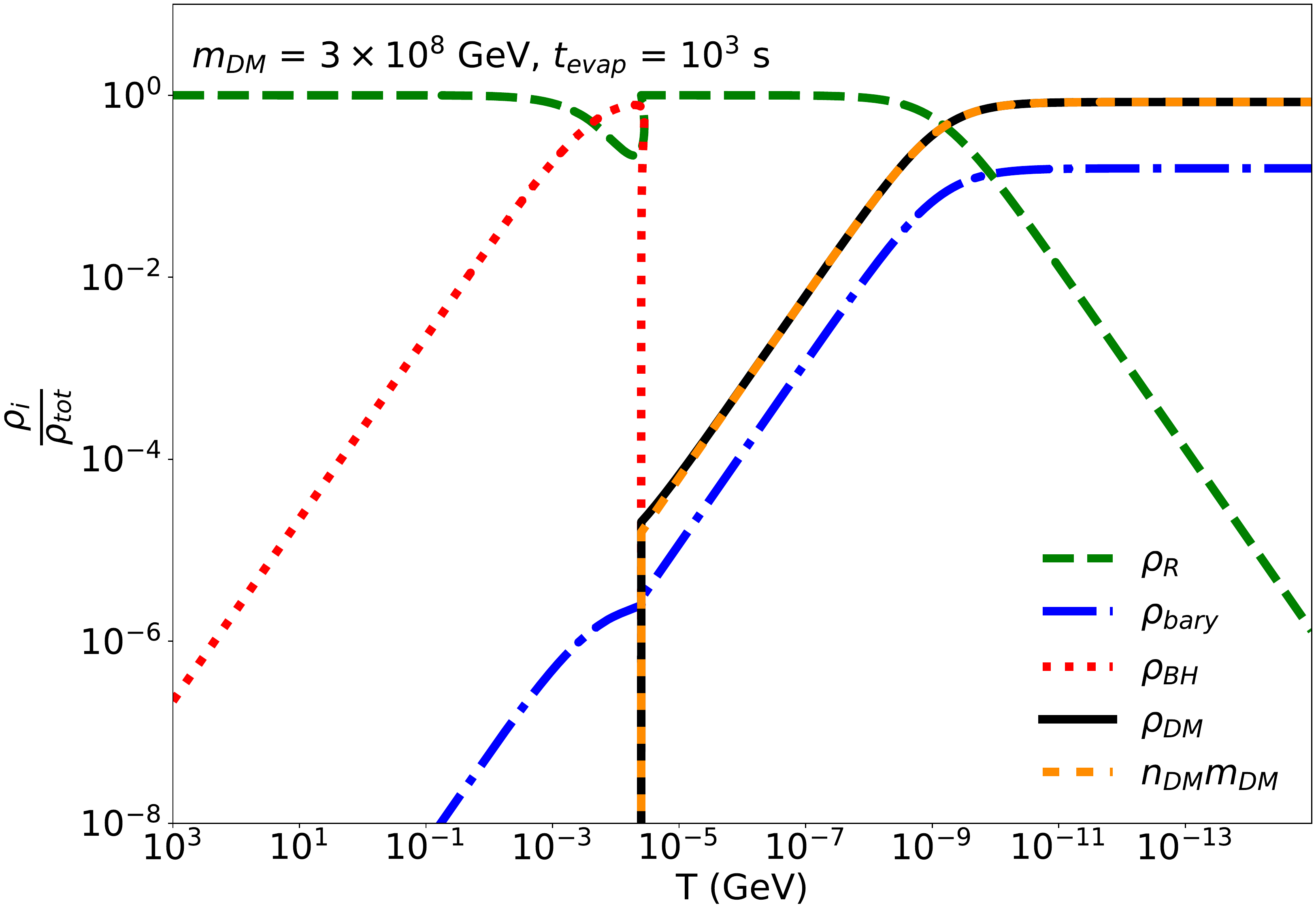} 
\caption{The evolution of the energy densities in Standard Model radiation, baryons, black holes, and dark matter, in a scenario in which the black holes evaporate with a lifetime of 1000 seconds almost entirely to dark matter particles (corresponding to $g_{\star, H}=10^6$ for $T_{\rm BH} \gg m_{\rm DM}$).  In each frame, the initial black hole abundance was chosen such that the Hawking radiation produces the entirety of the measured dark matter density. This corresponds to $\Omega_{\rm BH} = 3.7$ (upper left), $4.0 \times 10^2$ (upper right), $4.0 \times 10^3$ (lower left) and $1.2 \times 10^5$ (lower right). As we have throughout this paper, we define $\Omega_{\rm BH}$ as the value that would be the case today if the black holes had not evaporated.}
\label{evolDM1000}
\end{figure}

In the high-temperature limit ($T_{\rm BH} \gg m$), the energy emitted from a black hole in the form of a given particle species is equal to the mass loss rate in Eq.~\ref{loss}, for the appropriate choice of $g_{\star, H}$ (for example, $g_{\star, H}=4$ for a singlet Dirac fermion). Furthermore, the average energy of the radiated particles in this limit is equal to $\langle E\rangle = 3.15 \,T_{\rm BH}$ for the case of a fermion, and $\langle E\rangle =  2.70 \,T_{\rm BH}$ for a boson~\cite{Kolb:1990vq}. For lower values of $T_{\rm BH}$, the total energy and the total number of particles radiated are each suppressed. This suppression can be quantified by the following expressions for the total energy, and the total number of particles, radiated per unit time from a black hole in the form of particles of mass, $m$:
\begin{eqnarray}
F &\propto& \int^{\infty}_{m} \frac{(E^2-m^2)^{1/2}}{e^{m/T_{\rm BH}} \pm 1} E^2 dE \\
N &\propto& \int^{\infty}_{m} \frac{(E^2-m^2)^{1/2}}{e^{m/T_{\rm BH}} \pm 1} E dE, \nonumber
\end{eqnarray}
where the $\pm$ in the denominators apply to the case of fermions (+) and bosons (-), respectively. 

In practice, increasing the mass of the radiated hidden sector particles has the effect of delaying the ability of a given black hole to produce significant quantities of that particle species. In Figs.~\ref{evolDM} and~\ref{evolDM1000}, we plot the evolution of the energy densities in Standard Model radiation, baryons, black holes, and dark matter, in a scenario in which the black holes evaporate almost entirely to dark matter. More specifically, we adopt a total value of $g_{\star, H} = 10^6$ in the $T_{\rm BH} \gg m_{\rm DM}$ limit (of which all but $\simeq 108$ corresponds to Hawking radiation into dark matter particles with a common mass of $m_{\rm DM}$). In these two figures, we adopt $t_{\rm evap}=10$ and 1000 seconds, respectively, and in each frame we have selected a different value of $m_{\rm DM}$.\footnote{Although we consider only one value of $m_{\rm DM}$ at a time, one could also consider scenarios in which there is a spectrum of heavy hidden sector states.} In each case, we have set the the initial black hole abundance such that the Hawking radiation produces a final dark matter abundance that is equal to the total measured dark matter density. In these figures, we have plotted separately the total energy density of dark matter, $\rho_{\rm DM}$, and the number of density of these particles multiplied by their mass, $n_{\rm DM} \, m_{\rm DM}$. This distinction can be non-negligible, as the dark matter particles are not necessarily non-relativistic when they are initially radiated from a black hole. This is most noticeable in the case of $m_{\rm DM}=1$ TeV, which is not much larger than the initial temperature of the black holes under consideration.

\begin{figure}[t]
\includegraphics[width=0.49\textwidth]{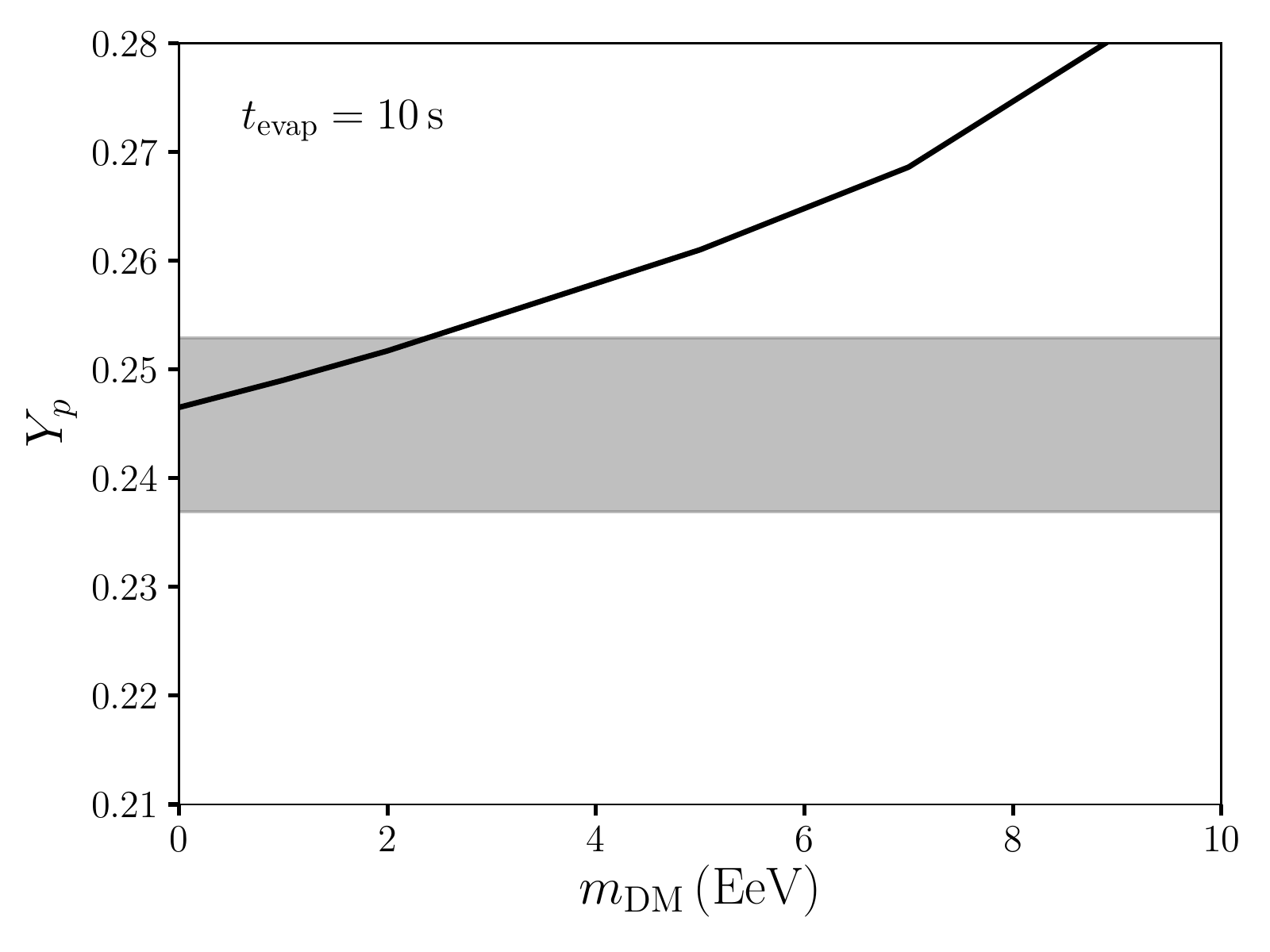} 
\includegraphics[width=0.49\textwidth]{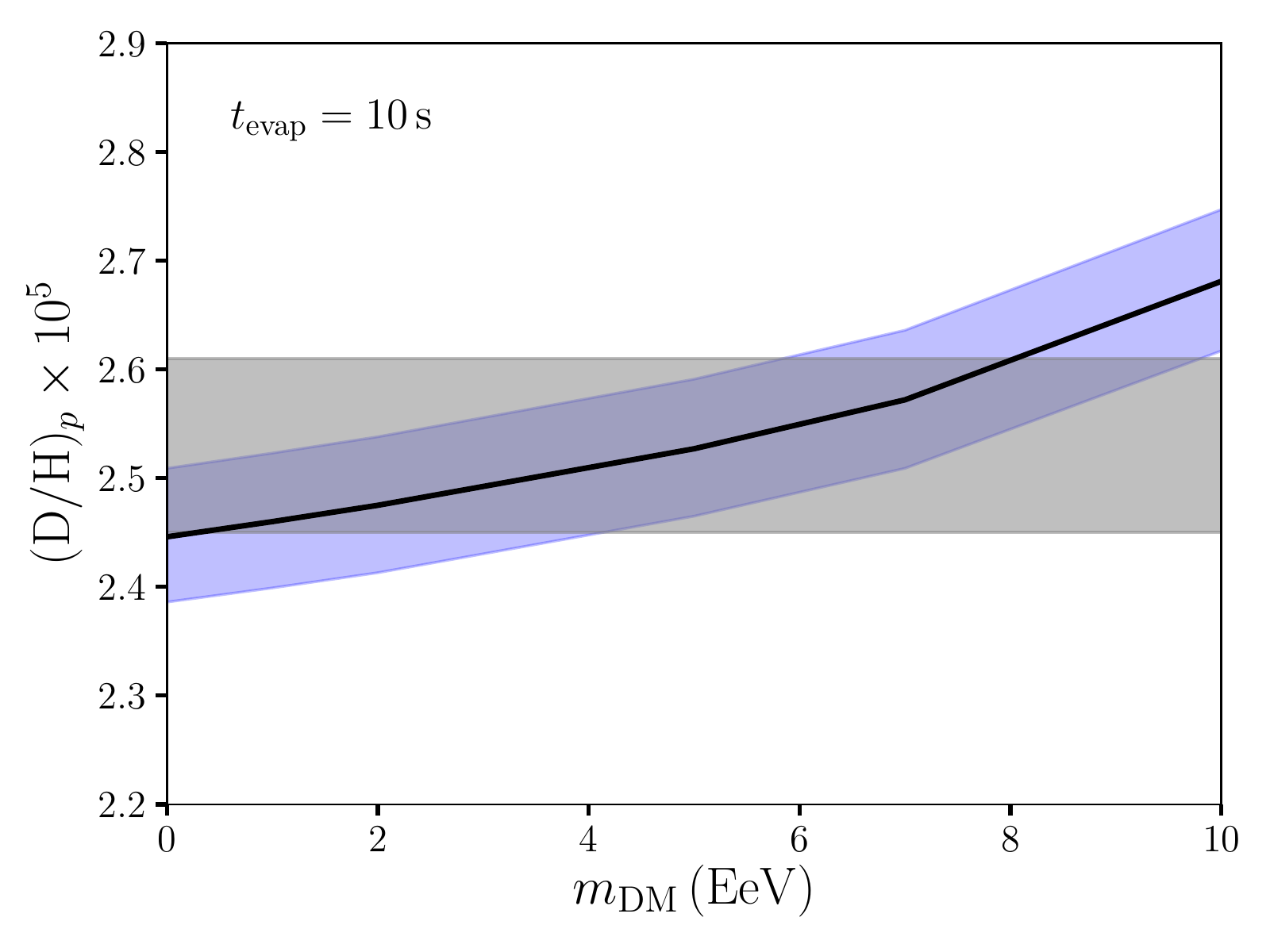} \\
\includegraphics[width=0.49\textwidth]{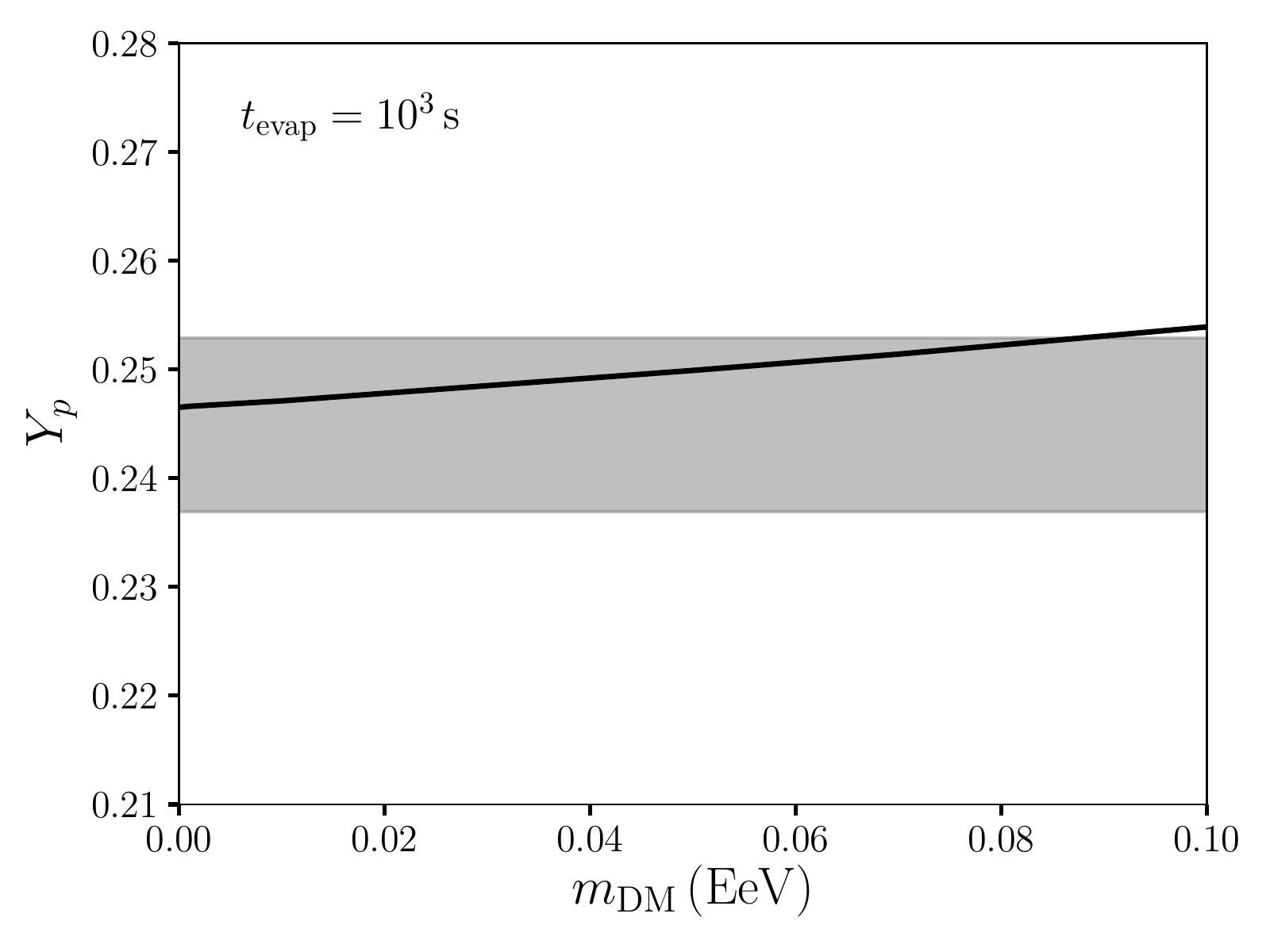} 
\includegraphics[width=0.49\textwidth]{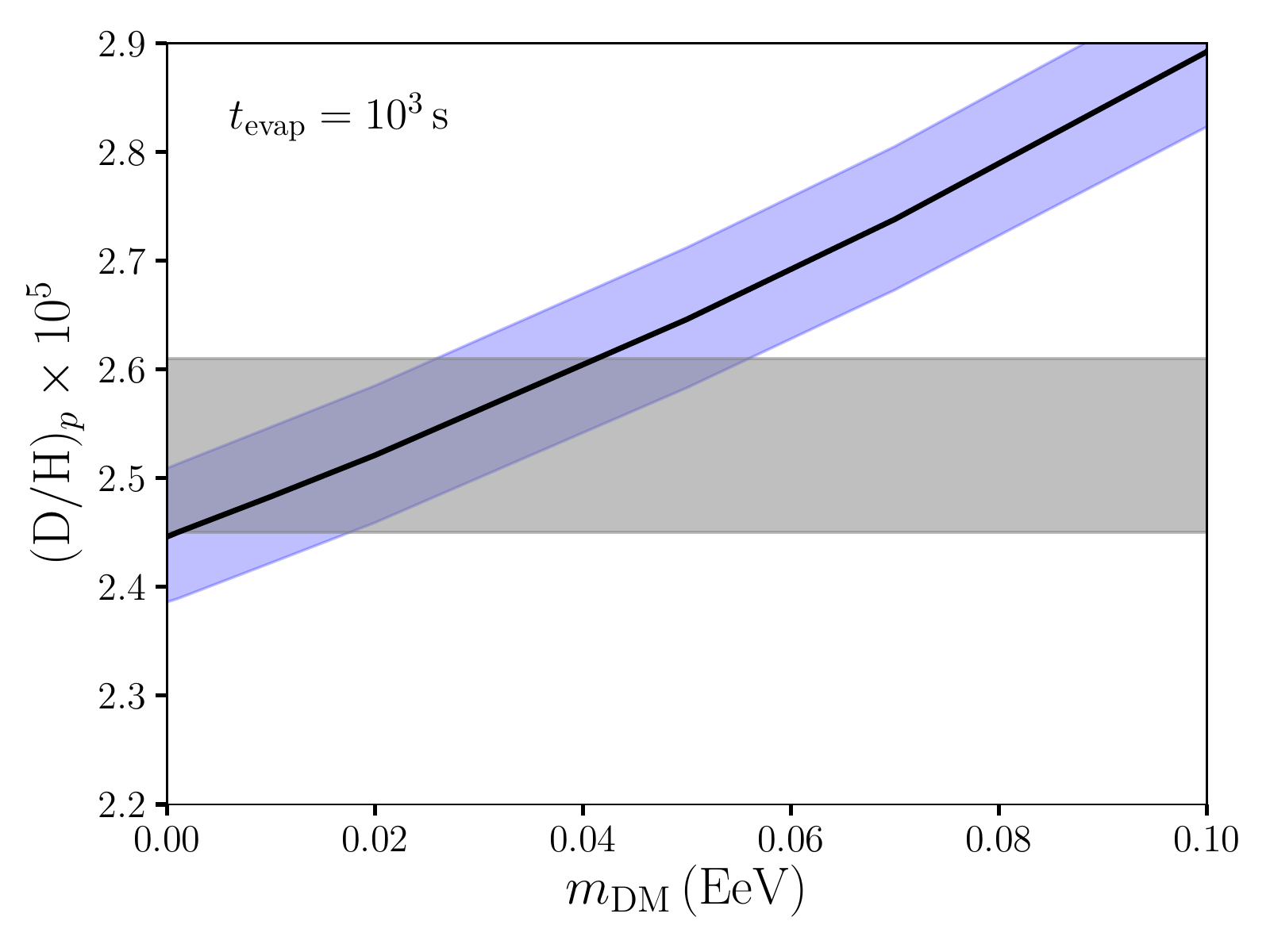} 
\caption{The impact on the primordial helium (left) and deuterium (right) abundances of black holes that evaporate largely to dark matter (corresponding to $g_{\star, H}=10^6$ for $T_{\rm BH} \gg m_{\rm DM}$). These results are given in terms of the dark matter particles' mass $m_{\rm DM}$, and in each case, the initial black hole abundance was chosen such that the Hawking radiation produces the entirety of the measured dark matter density. The grey bands represent the measured values (at $2\sigma$), while the blue band in the right frame denotes the systematic uncertainty associated with the nuclear reaction rates (as described in Sec.~\ref{BSM}).}
\label{darkmatter}
\end{figure}

 In Fig.~\ref{darkmatter}, we show how these scenarios impact the primordial helium and deuterium abundances, focusing on the effects of the black holes and their evaporation products on the expansion rate. Although we show these results in terms of $m_{\rm DM}$, they can be directly translated into values of the black hole abundance, $\beta'$ or $MY$. In the $t_{\rm evap} = 10 \, {\rm s}$ case, the expansion rate can be significantly altered during the time of proton-neutron freeze-out, enhancing the neutron abundance at early times and leading to constraints based on the measured helium mass fraction, $Y_p$. For this lifetime, the measured value of $Y_p$ allows us to constrain $\beta' \lsim 2 \times 10^{-15}$. In the $t_{\rm evap} = 10^3 \, {\rm s}$ case, the measured deuterium abundance instead provides the most stringent constraint, allowing us to constrain $\beta' \lsim 5 \times 10^{-16}$. Additionally in this case, if the black hole abundance is large, the baryon abundance will be enhanced at early times (as can be seen in the lower right frame of Fig.~\ref{evolDM1000}), impacting the rates of fusion and potentially ruining the successful prediction of $Y_p$.
 
When comparing these results to those presented in Fig.~\ref{mainSM}, we reach the following conclusions. First, the black holes and their dark matter Hawking evaporation products only observably impact the expansion history of the universe in regions of parameter space that are already ruled out as a consequence of Hawking evaporation into Standard Model particles.  It is entirely possible, however, that such black holes could generate the entirety of the observed dark matter abundance. For the case of $t_{\rm evap} \sim 10 \, {\rm s}$ with $g_{\star, H} \gg 10^2$, this can be self-consistently attained so long as $m_{\rm DM} \lsim (10^6 \, {\rm GeV}) \times (g_{\star, H}/10^4)$. For $g_{\star, H} \sim 10^2$, we instead find that we must require $m_{\rm DM} \lsim (10^6 \, {\rm GeV}) / g^{\rm DM}_{\star, H}$ in order to obtain the observed dark matter abundance, where $g^{\rm DM}_{\star, H}$ is the contribution of the dark matter species to $g_{\star, H}$. For heavier dark matter candidates, it is not possible to produce the total measured abundance without violating the constraints presented in this study (unless $t_{\rm evap} \lsim 10 \, {\rm s}$). In the case of $t_{\rm evap} \sim 10^3 \, {\rm s}$, these requirements are more stringent. In particular, to obtain the full measured dark matter abundance from such black holes, we must require $m_{\rm DM} \lsim (10 \, {\rm GeV}) \times (g_{\star, H}/10^4)$ (for $g_{\star, H} \gg 10^2$) or  $m_{\rm DM} \lsim (10 \, {\rm GeV}) / g^{\rm DM}_{\star, H}$ (for $g_{\star, H} \sim 10^2$).

Compared to dark matter candidates that are produced as a WIMP-like thermal relic, particles generated through Hawking radiation are much more energetic, raising the question of whether they would behave as cold dark matter (as opposed to warm or hot dark matter)~\cite{Baldes:2020nuv}. In the $m_{\rm DM} \gg T_i$ limit, we find the average energy of a radiated dark matter particle by integrating from the time at which $T_{\rm BH} \sim m_{\rm DM}$ to the end of a black hole's evaporation, resulting in $\langle E_{\rm DM} \rangle \sim 6 \, m_{\rm DM}$. By then relating $\langle E_{\rm DM }\rangle \sim 3 T_{\rm DM}$, we can estimate the approximate free-streaming length~\cite{Kolb:1990vq}:
\begin{eqnarray}
\lambda_{\rm fs} &=& \int^{t_{\rm nr}}_0 \frac{dt}{a(t)} \approx 1\,{\rm Mpc} \times \bigg(\frac{T_{\rm DM}}{T}\bigg) \, \bigg(\frac{0.3 \, {\rm keV}}{m_{\rm DM}}\bigg) \\
&\sim& 6 \times 10^{-4} \, {\rm Mpc} \times  \bigg(\frac{t_{\rm evap}}{{\rm s}}\bigg)^{0.5}. \nonumber
\end{eqnarray}
From this estimate, it follows that any stable, feebly-interacting particles heavier than $T_i$ that are generated via Hawking radiation will act as cold dark matter ($\lambda_{\rm fs} \lsim {\rm Mpc}$) so long as $t_{\rm evap} \lsim 3 \times 10^{6}\, {\rm s}$. 

Considering next the case of $m_{\rm DM} \ll T_i$, we estimate $T_{\rm DM} \sim T_i$, which leads to the following:
\begin{eqnarray}
\lambda_{\rm fs} &=& \int^{t_{\rm nr}}_0 \frac{dt}{a(t)} \approx 1\,{\rm Mpc} \times \bigg(\frac{T_{\rm DM}}{T}\bigg) \, \bigg(\frac{0.3 \, {\rm keV}}{m_{\rm DM}}\bigg) \\
&\sim& 3 \times 10^{-2} \, {\rm Mpc} \times  \bigg(\frac{T_i/m_{\rm DM}}{100}\bigg) \, \bigg(\frac{t_{\rm evap}}{{\rm s}}\bigg)^{0.5}. \nonumber
\end{eqnarray}
Thus any stable, feebly-interacting particles lighter than $T_i$ that are generated via Hawking radiation will act as cold dark matter ($\lambda_{\rm fs} \lsim {\rm Mpc}$) so long as $t_{\rm evap} \lsim 10^{7}\, {\rm s} \times (m_{\rm DM}/T_i)^2$. 

To summarize the results of this subsection, for black holes with $t_{\rm evap} \sim 10 \, {\rm s}$, there exist a wide range of scenarios in which the dark matter could be produced through Hawking evaporation, especially if $m_{\rm DM} \sim {\rm GeV}-{\rm PeV}$ (a wider range of $m_{\rm DM}$ are also possible, but only if $g_{\star, H}$ is very large). For black holes with $t_{\rm evap} \sim 10^3 \, {\rm s}$, the range of such possibilities are more restricted, but are viable if $m_{\rm DM} \sim 10\,{\rm GeV}$ (or for a wider range of $m_{\rm DM}$ if $g_{\star, H}$ is very large).

\subsection{TeV-Scale Supersymmetry}

As another example, we will consider a scenario in which most of the superpartners of the Standard Model (in particular, the squarks and gluinos) are not much heavier than the current constraints placed by the Large Hadron Collider (LHC), $m_{\rm SUSY} \sim 2$ TeV~\cite{Aaboud:2017vwy,Sirunyan:2017cwe}. In this case, black holes heavier than $M\sim 5 \times 10^9$ g (corresponding to $T_{\rm BH}\sim 2$ TeV) still evaporate almost entirely into Standard Model particles. But as the mass of a black hole falls below this threshold, it will begin to also evaporate into the full spectrum superpartner particles. Numerically, this has the effect of changing $g_{\star, H}$ from Standard Model value of 108, to approximately 316 (for the case of the Minimal Supersymmetric Standard Model, MSSM).\footnote{By supersymmetrizing the Standard Model, the value of $g_{\star, H}$ does not merely double, but is further enhanced as a result of the lower spins of most of the sparticle degrees-of-freedom.} So whereas a black hole with a mass of $M\sim 5 \times 10^9$ g will evaporate in $t_{\rm evap} \sim 50$ s, assuming Standard Model particle content, this instead occurs in $t_{\rm evap} \sim 17$ s in the presence of low-energy supersymmetry. This has the effect of relaxing the constraints on black holes in this mass range by a factor of $\sim2-3$.

Additionally, if R-parity is conserved, each superpartner radiated from a black hole will ultimately decay to a lightest supersymmetric particle (LSP). If the LSP is relatively heavy, such as  a neutralino, it could serve as a candidate for dark matter. On the other hand, a very light LSP (perhaps in the form of a gravitino or axino) could act as dark radiation in this context. If $m_{\rm LSP} \ll m_{\rm SUSY}$, this will have little impact on the resulting constraints. If the sparticle spectrum is highly compressed ($m_{\rm LSP} \sim m_{\rm SUSY}$), however, the majority of the energy in the Hawking radiation will be in the form of LSP dark matter, reducing the potential to break up helium (and produce deuterium) by a factor of up to $\sim2-3$. 


The abundance of LSP dark matter that is generated through Hawking radiation in this scenario is given by:
\begin{eqnarray}
\Omega_{\rm LSP} &\simeq& \Omega_{\rm BH} \times f_{\rm SUSY} \times \frac{m_{\rm LSP}}{m_{\rm SUSY}} \\
&\sim& \frac{\beta'}{2.2 \times 10^{-20}} \, \bigg(\frac{10^{10}\,{\rm g}}{M}\bigg)^{0.5} \times f_{\rm SUSY} \times \frac{m_{\rm LSP}}{m_{\rm SUSY}}, \nonumber
\end{eqnarray}
where $\Omega_{\rm BH}$ is defined as the value of $\rho_{\rm BH}/\rho_{\rm crit}$ that would be the case today if the black holes had not evaporated (see Eq.~\ref{omegabh}), and $f_{\rm SUSY}$ is the fraction of energy in Hawking radiation that is in the form of superparticles. For $M \lsim 5 \times 10^9 \, {\rm g} \times (2 \, {\rm TeV}/m_{\rm SUSY})$, the black hole can efficiently radiate superpartners throughout its evaporation, and this fraction is simply given by: $f_{\rm SUSY} \sim g^{\rm SUSY}_{\star, H}/g_{\star, H}$ where $g^{\rm SUSY}_{\star, H} \simeq 208$. For more massive black holes, 
\begin{eqnarray}
f_{\rm SUSY} \sim \bigg(\frac{5 \times 10^9 \, {\rm g}}{M}\bigg) \, \bigg(\frac{2 \, {\rm TeV}}{m_{\rm SUSY}}\bigg) \, \bigg(\frac{g^{\rm SUSY}_{\star, H}}{g_{\star, H}}\bigg).
\end{eqnarray}
For example, for $m_{\rm SUSY} = 2 \, {\rm TeV}$, $m_{\rm LSP} = 1 \, {\rm TeV}$, and $M=5 \times 10^9 \, {\rm g}$, the value of $\Omega_{\rm LSP}$ is equal to the measured dark matter density for an initial black hole abundance of $\beta' \simeq 10^{-20}$. Given the more rapid rate of evaporation due to superpartner production, such a scenario is compatible with the measured light element abundances.


%

Phenomenology similar to that described in this subsection could also arise within the context of other weak-scale extensions of the Standard Model, such as mirror models or Twin Higgs models. Such models are motivated by the little hierarchy problem~\cite{Chacko:2005pe,Chacko:2018vss}, and generally include a copy of some or all of the Standard Model particle content, with masses rescaled by the larger vacuum expectation value of the mirror Higgs boson (for a review, see Ref.~\cite{Berezhiani:2005ek}). In these models, the lightest mirror-charged state is typically stable~\cite{Chacko:2016hvu,Chacko:2018vss}, opening up the possibility that a stable population of such particles could be generated through the process of Hawking evaporation.

\section{Summary and Conclusion}

It is plausible that the early universe contained a substantial population of black holes. Any such objects light enough to evaporate prior to the onset of BBN are largely unconstrained by current observations. In contrast, black holes that evaporated during or shortly after the BBN era can be constrained by measurements of the primordial light element abundances. In this study, we revisited the impact of evaporating black holes on BBN, updating the relevant measurements and expanding the discussion to include cases in which black holes can evaporate into particles beyond the Standard Model. 

Recent improvements in the determination of the primordial deuterium abundance have made it possible to significantly strengthen the constraints on primordial black holes relative to those presented in previous work. Our main results are shown in Fig.~\ref{mainSM}, where (assuming only Standard Model particle content) we summarize our constraints on the initial abundance of primordial black holes which evaporate on a timescale of $10^{-1}$ to $10^{13} \, {\rm s}$ (corresponding to a mass range of $\sim 6\times 10^8$ to $\sim 2\times 10^{13}$ grams). For $t_{\rm evap} \lsim 80 \, {\rm s}$, these constraints are largely the consequence of the neutron abundance at early times, which is sensitive to the expansion rate at the time of proton-neutron freeze-out, as well as to the presence of energetic mesons which can convert protons into neutrons (and vice versa). For longer-lived black holes, the constraints are instead dominated by the hadrodissociation and photodissociation of helium nuclei, each of which can significantly increase the observed abundance of primordial deuterium. 

Whereas previous papers studying the impact of primordial black holes on BBN have focused on Hawking evaporation into Standard Model particles, we have extended this discussion to include scenarios beyond the Standard Model. Given the purely gravitational nature of Hawking evaporation, black holes produce all particle species lighter than the black hole's temperature, regardless of their charges or couplings. As a consequence, the rate at which black holes evaporate and the types of particles that are produced through this process depend on the complete particle spectrum, including any and all such species that might exist beyond the confines of the Standard Model. From this perspective, it is particularly interesting to consider scenarios that feature large numbers of feebly-coupled degrees-of-freedom. In exploring such hidden sector models, we have considered constraints on light, stable products of Hawking radiation (which act as dark radiation), as well as massive stable particles (which act as dark matter). For $t_{\rm evap} \lsim 10^2 \, {\rm s}$, we have placed constraints on black holes with light hidden sectors that are comparably stringent to those derived from measurements of the CMB (longer-lived black holes are more strongly restricted by the CMB). For relatively short-lived black holes ($t_{\rm evap} \lsim 10^3 \, {\rm s}$), we have identified a wide range of scenarios in which the entirety of the dark matter abundance could be produced through Hawking evaporation, especially if $m_{\rm DM} \sim {\rm GeV}-{\rm PeV}$ (other values of $m_{\rm DM}$ are also possible, but only if $g_{\star, H}$ is very large or $t_{\rm evap} \lsim 10^{-1}\, {\rm s}$). For longer-lived black holes, the combined constraints from the measured deuterium abundance and large scale structure (as related to the dark matter's free-streaming length) are more restrictive. We also consider evaporating black holes within the context of TeV-scale supersymmetry, finding a non-negligible impact on the resulting constraints, and identifying scenarios in which Hawking evaporation could produce an abundance of neutralinos (or other LSPs) that is in good agreement with the measured dark matter density.

To end on a more general note, the epoch of BBN provides us with critical information pertaining to the energy content of the universe at early times. Measurements of the primordial element abundances serve as a window into this period, enabling us to test and constrain a wide range of possible new phenomena, including that of primordial black holes. 

\acknowledgments

This work has been supported by the National Science Foundation Graduate Research Fellowship Program under Grant No. DGE-1746045, and by the Fermi Research Alliance, LLC under Contract No. DE-AC02-07CH11359 with the U.S. Department of Energy, Office of High Energy Physics. Any opinions, findings, conclusions, or recommendations expressed in this material are those of the authors and do not necessarily reflect the views of the National Science Foundation or Department of Energy.

\bibliography{PBH}

\end{document}